\documentclass[preprint,11pt,3p]{elsarticle}

\usepackage{graphicx}
\usepackage{amssymb}
\usepackage{amsmath}
\usepackage{amsthm}
\usepackage{gensymb}

\usepackage[utf8]{inputenc}
\usepackage[T1]{fontenc}
\usepackage{float}
\usepackage{booktabs}
\usepackage{colortbl}
\usepackage{pbox}
\usepackage{pifont}
\usepackage{setspace}
\usepackage{multirow}
\usepackage{lineno}
\usepackage{afterpage}
\usepackage[caption=false,font=footnotesize,labelfont=sf,textfont=sf]{subfig}
\usepackage[prependcaption, textsize=tiny, linecolor=red, backgroundcolor=red!25, bordercolor=red, textwidth=3.1cm]{todonotes}
\usepackage[plainpages=false, bookmarks=false, hidelinks]{hyperref}        

\hypersetup{
  pdftitle={Global Fire Season Severity Analysis and Forecasting}, 
  pdfauthor={Leonardo N Ferreira},
  pdfcreator={Leonardo N Ferreira} 
  colorlinks=true,          
  linkcolor=blue,           
  citecolor=blue,           
  filecolor=magenta,          
  urlcolor=blue,
  bookmarksdepth=2
}

\newcommand{\xmark}{\ding{55}} 

\makeatletter
\def\ps@pprintTitle{%
   \let\@oddhead\@empty
   \let\@evenhead\@empty
   \let\@oddfoot\@empty
   \let\@evenfoot\@oddfoot
}
\makeatother

\begin{document}

\begin{frontmatter}


\title{The Effect of Time Series Distance Functions on\\Functional Climate Networks}




\author[address1,address2,address3]{Leonardo N. Ferreira\corref{mycorrespondingauthor}}
\ead{ferreira@leonardonascimento.com}
\cortext[mycorrespondingauthor]{Corresponding author}

\author[address4]{Nicole C. R. Ferreira}
\ead{nicole.resende@yahoo.com.br}

\author[address5]{Elbert E. N. Macau}
\ead{elbert.macau@unifesp.br}

\author[address3,address6]{Reik V. Donner}
\ead{reik.donner@h2.de}

\address[address1]{Associated Laboratory for Computing and Applied Mathematics, National Institute for Space Research,\\Av. dos Astronautas, 1758, Jardim da Granja, CEP: 12227-010. S\~ao Jos\'{e} Dos Campos - SP, Brazil}

\address[address2]{Department of Physics, Humboldt University, Newtonstra{\ss}e 15, 12489 Berlin, Germany}

\address[address3]{Research Department IV -- Complexity Science, Potsdam Institute for Climate Impact Research (PIK)\\Member of the Leibniz Association, Telegrafenberg A56, 14473 Potsdam, Germany}

\address[address4]{Center for Weather Forecast and Climatic Studies, National Institute for Space Research,\\Rodovia Presidente Dutra, Km 40, CEP: 12630-000. Cachoeira Paulista - SP, Brazil.}

\address[address5]{Institute of Science and Technology, Federal University of S\~ao Paulo,\\Av. Cesare Monsueto Giulio Lattes, 1201 - Eugênio de Melo, São José dos Campos - SP Brazil}

\address[address6]{Department of Water, Environment, Construction and Safety, Magdeburg-–Stendal University of Applied Sciences, Breitscheidstr. 2, 39114 Magdeburg, Germany}



\begin{abstract}
Complex network theory provides an important tool for the analysis of complex systems such as the Earth's climate. In this context, functional climate networks can be constructed using a spatiotemporal climate dataset and a suitable time series distance function. The resulting coarse-grained view on climate variability consists of representing distinct areas on the globe (i.e., grid cells) by nodes and connecting pairs of nodes that present similar time series. One fundamental concern when constructing such a functional climate network is the definition of a metric that captures the mutual similarity between time series. Here we study systematically the effect of 29 time series distance functions on functional climate network construction based on global temperature data. We observe that the distance functions previously used in the literature commonly generate very similar networks while alternative ones result in rather distinct network structures and reveal different long-distance connection patterns. These patterns are highly important for the study of climate dynamics since they generally represent pathways for the long-distance transportation of energy and can be used to forecast climate variability on subseasonal to interannual or even decadal scales. Therefore, we propose the measures studied here as alternatives for the analysis of climate variability and to further exploit their complementary capability of capturing different aspects of the underlying dynamics that may help gaining a more holistic empirical understanding of the global climate system.
\end{abstract}




\end{frontmatter}


\section{Introduction}
\label{sec_introction}

Statistical climatology provides a great variety of tools for studying the dynamics of meteorological variables across time and space. As a recent approach complementing more traditional analysis techniques, during the last decades complex network tools have emerged as an alternative to statistically analyze and model complex systems such as the Earth's climate \cite{Dijkstra2019,Donner2017,boers14,donges09,jingfang17,steinhaeuser11,tsonis04,ferreira20}. In this context, the resulting networks are called functional climate networks and can be constructed in different ways. The most common procedure consists of dividing a spatiotemporal climate dataset into grid cells. In this case, network nodes represent the cells of the grid and links are established if the time series of climate variability associated with different cells fulfil a certain criterion of statistical similarity. The main goal of such a climate network description is to highlight the essential interrelationships among the spatially distributed grid points that govern certain relevant aspects of climate variability regionally or even globally.
 
One question that has not been systematically addressed so far in the context of constructing functional climate networks concerns the choice of a metric that captures similarities between two time series. The majority of previous works has used linear statistical association measures like covariance or Pearson correlation coefficient with or without time-lag \cite{donges09,jingfang17,tsonis04,guez14,tsonis08}. With these measures, it has been possible to construct climate networks and detect interesting climatological patterns, while they have been conceptually restricted to statistical similarities captured by the concept of correlation or straightforward nonlinear generalizations thereof like mutual information \cite{donges09,palus18}. However, there exist a plethora of other distance functions that could be equally applied to the construction of climate networks \cite{cha07,deza09,esling12,ferreira16}. For simplicity reasons, we use here the term ``distance function'' to generally refer to these measures even though some of them are no distance functions in the strict mathematical sense. Some of these functions can capture statistical similarities in climatological variables that correlations cannot detect. Thus, the central question here is whether those alternative distance functions can generate climate networks that highlight different climatological phenomena than the ``classical'' ones and consequently reveal other meaningful patterns.

In this paper, we evaluate how different choices of time series distance functions affect the properties of the resulting functional climate networks. We are particularly interested in the topological differences among the resulting networks and in understanding the meaning of the climatological patterns captured by them. We use 29 different distance functions and apply complex network theory \cite{barabasi16,boccaletti06,costa07} to analyze the networks constructed based on global near-surface air temperature data. In the construction process, we employ a parameter $p$ that controls the number of edges (i.e., the edge density) for all the networks. This procedure generates networks with the same number of edges. When the edge density is low ($p \to 0$), the resulting networks have only a few edges connecting the most similar time series, while those networks tend to exhibit disconnected components. Conversely, high edge densities ($p \to 1$) generate densely connected networks, but many of the resulting edges connect pairs of time series that are not quite similar. Choosing an intermediate value of $p$ is hence desirable, since it preserves only those edges that represent the most similar time series and avoids the creation of strongly fragmented networks \cite{Wiedermann2017}.

We start by studying the influence of the edge density $p$ on the networks. For the considered network size (i.e., number of grid points), we observe a transition phase in the interval $0.0004 \leq p \leq 0.1$ from disconnected components to connected networks. In this interval, the resulting networks connect only the most similar time series while the networks are still not highly fragmented. In this range of edge densities, the resulting climate networks exhibit a (likely geometrically induced \cite{Bialonski2010,Bialonski2011,Hlinka2012,Wiedermann2016}) small-world effect along with degree distributions that decay faster than a power law. Regarding the network topology, we find that the correlation, cross-correlation, mutual information and maximal information coefficient generate similar networks. In contrast, alternative distance functions generate very different network structures. We also observe variations in the connectivity among geographically distant nodes (teleconnections) which we compare with those previously observed in the literature. These long-distance edges, like such emanating from the El Niño--Southern Oscillation (ENSO) region in the eastern tropical Pacific, are important because they represent the transport of energy and atmospheric wave propagation on global scales \cite{palus11,nigam15,dong15}. Such teleconnections are also practically relevant due to their capability to anticipate climatological events and enhanced forecast skill from subseasonal up to even decadal time scales \cite{alexander02,boers19}. Therefore, our results are significant for climate sciences since the identified teleconnection patterns capture distinct types of underlying dynamics. We here show that alternative distance functions permit the study of climate variability from complementary perspectives and could thereby contribute to a better understanding of the climate system. We present these results in details in the following sections.

We organize the remainder of this paper as follows. In Section~\ref{sec:materials_methods}, we introduce the methodological framework employed in this paper. We summarize the relevant concepts of functional climate network construction and time series distance functions, along with the data and specific methodology used to obtain our results. In Section~\ref{sec:results}, the results of our study are presented in three parts focusing on global network features, network similarities and teleconnection patterns, respectively. Finally, in Section~\ref{sec:conclusion}, we present some conclusions drawn from our reported findings.

\section{Materials and methods}
\label{sec:materials_methods}

\subsection{Functional climate networks}

\begin{figure}[!ht]
  \centering
  \includegraphics[width=1.0\linewidth]{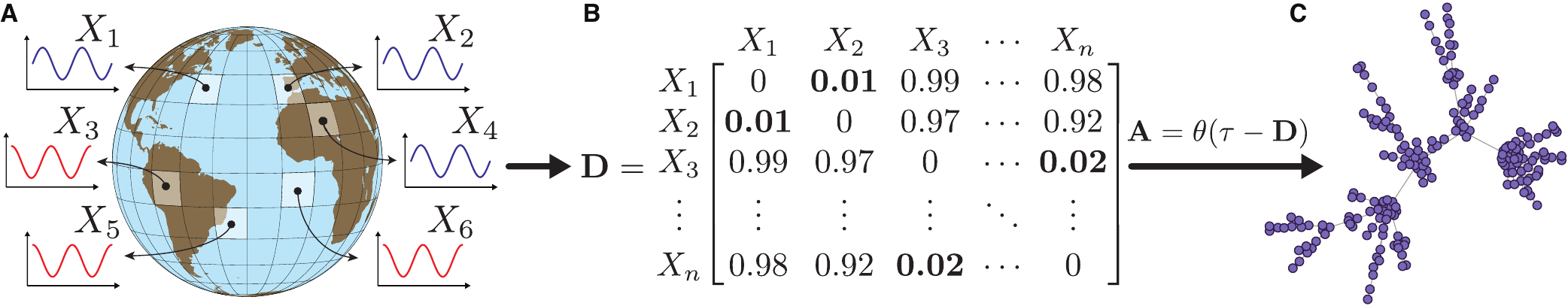} 
  \caption{Schematic illustration of the process of climate network construction. (A) The first step consists of building a spatiotemporal gridded climate dataset. This dataset can be constructed by dividing a region of interest into grid cells that represent climate variability in smaller regions in terms of, e.g., air temperature, sea level pressure, precipitable water or wind speed. (B) Second, a time series distance function is used to construct a distance matrix $\mathbf{D}$ whose elements $D_{ij}$ provide the distance between two time series $X_i$ and $X_j$. (C) The distance matrix $\mathbf{D}$ is transformed into an adjacency matrix $\mathbf{A}$ by taking all values from $\mathbf{D}$ smaller than a selected threshold value $\tau$, i.e., $\mathbf{A} = \Theta(\tau - \mathbf{D})$, where $\Theta(\bullet)$ is the Heaviside step function. High threshold values ($\tau \to 1$ in case of normalized distance functions) generate densely connected networks while very low ones ($\tau \to 0$) generate fragmented networks with disconnected components.}
  \label{fig:net_contruction} 
\end{figure}

Complx network theory has been applied for studying many different types of complex systems. In the context of the Earth's climate system, networks can be used to represent the essential information provided by a spatiotemporal gridded climate dataset. In functional climate networks, every grid cell from a given dataset corresponds to a node while only the most similar pairs of nodes are connected in the network (Fig.~\ref{fig:net_contruction}). This approach raises two natural questions: How to measure similarity between time series and which threshold value for this measure to use for identifying two time series as similar and, hence, connected? Both choices have a direct effect on the resulting network topology. The majority of the existing literature \cite{donges09,tsonis04,guez14,tsonis08,yamasaki08} has used the absolute value of the Pearson correlation coefficient (PCC), either as instantaneous or lagged correlation, thereby integrating information on both, strong positive and strong negative correlations.

The first application of network science in climate studies was presented by Tsonis \emph{et~al.}~\cite{tsonis04}. In that work, the authors used wind flow data from the NCEP/NCAR reanalysis dataset \cite{kalnay96} and the PCC ($>0.5$) at lag zero to measure similarity between time series. They observed that the resulting network presented an apparent small-world property and was divided into two subgraphs, one of them representing the tropics and the other the extratropical regions. In the tropics, most nodes were densely interconnected, whereas the higher latitudes presented only some distinct highly connected nodes (hubs) and possessed characteristics resembling a scale-free network. The conclusion of the authors at that point was that the subnetwork corresponding to the equatorial region acts as an agent connecting the two hemispheres and thereby allowing for information to travel between them.  


Yamasaki \emph{et~al.}~\cite{yamasaki08} used globally distributed temperature time series and observed that the presence of El Niño conditions in the eastern tropical Pacific significantly affects the structure of the resulting network. The authors showed that while El Niño does not necessarily change the temperature in different parts of the world significantly, many network edges are broken under its influence, so that the remaining ones serve as a measure for characterizing this phenomenon. Their tentative conclusion was that the network connections encompass information that had not been captured by previously employed more traditional statistical analyses, which suggests that complex network theory can reveal patterns not observed before by established techniques from statistical climatology. Several later studies have reinforced this idea. For example, Berezin \emph{et~al.}~\cite{berezin12} used temperature and geopotential height time series to construct functional climate networks and observed that the relationship between their fluctuations in different geographical regions presents a robust network pattern, while the common belief prior to their study was rather that these variations were not stable and difficult to predict. Steinhaeuser \emph{et~al.}~\cite{steinhaeuser11} showed that the attributes extracted from communities in climate networks can be used as predictive climatic indices and present statistically better results than traditional clustering methods. Boers \emph{et~al.}~\cite{boers14} demonstrated the possibility to predict climatic extremes using complex networks. Functional climate networks have also been used to investigate large-scale circulation patterns, climate variability modes, their teleconnections, and for model intercomparison \cite{tsonis08,yamasaki08,deza15,bracco18}.

The majority of the aforementioned studies has employed correlation or covariance as distance functions. This implies that these studies have been limited to similarities captured by correlations among the variables under study. In this paper, we show that other distance functions can also be used for the construction of climate networks, an idea that was similarly proposed earlier by Pelan \emph{et~al.}~\cite{pelan11}. In the latter work, the authors used six different distance functions and analyzed the effect of edge density on the topology and the clustering predictive capability of the networks. Here, we thoroughly expand this setting by using in total 29 different distance functions and examining the topological differences between the resulting networks.

\subsection{Time series distance functions}
\label{subsec:ts_dist}

A time series distance function is a measure of how different two time series are. For the sake of brevity, the term ``distance function'' will be used here to refer to a time series distance function in the following. Other types of distance functions, e.g., network distance functions, will be referred to with their complete terms in the text. A plethora of distance functions have been proposed in the literature, which can be broadly classified into four main categories \cite{esling12}: shape-based, edit-based, feature-based and structure-based. 

In the following, we provide an overview on some time series distance function concepts that could be used to generate climate networks, without attempting to be complete in this task. Notably, we only consider measures that make use of all time series values, thereby excluding event synchrony measures like event synchronization \cite{boers14,boers19} or event coincidence analysis \cite{Wolf2020} that have been widely used for studying spatiotemporal patterns of extreme climate events from a complex network perspective. Moreover, we also discard measures of phase synchronization \cite{Yamasaki2009} as well as such based on the physical concept of causality and its statistical implementation in terms of coupled stochastic process models or information theory (e.g., Granger causality strength, transfer entropy, and similar measures) from the presented analysis.

\paragraph{Shape-based distance functions.} 

\begin{figure}[ht]
  \centering
  \includegraphics[width=.7\linewidth]{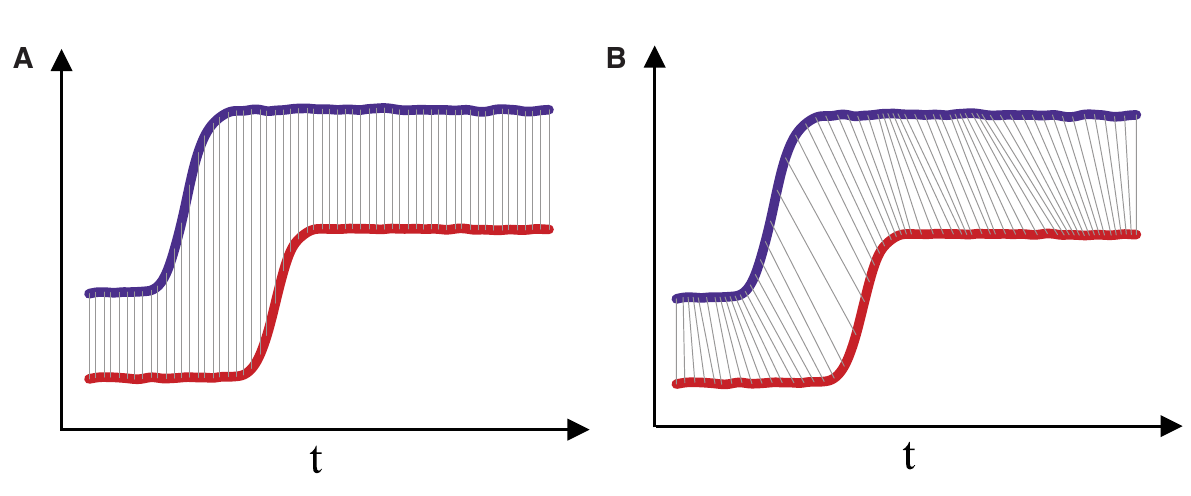}
  \caption{Time series comparison using lock-step and elastic measures. The total distance is proportional to the integral length of the gray lines. (A) Lock-step measures compare fixed (one-to-one) pairs of elements. (B) Elastic measures perform an alignment of the series and allow one-to-many comparisons of elements.}
  \label{fig:align} 
  \end{figure}

This kind of measures compare the overall shape of two time series. The most famous measures are $L_p$ distances, defined as

  \begin{equation}
    \label{eq:lp_dist}
     d_{L_p}(X,Y) = \Bigg( \sum_{i=1}^{t}(X_i - Y_i)^p\Bigg)^{\frac{1}{p}}.
  \end{equation}
\noindent 
When $p=2$, the $L_2$ distance is the well-known Euclidean distance. Other common choices are the $L_1$ and $L_\infty$ distances that correspond to the Manhattan and Chebyshev distances, respectively. The $L_p$ measures are parameter-free and intuitive but fail to capture similarity in many cases, for example, when two time series have similar shapes but are not aligned. This problem is illustrated in Fig~\ref{fig:align}. $L_p$ measures fail in this case because they compare values at fixed positions in the series; they are therefore also called \emph{lock-step measures}. To solve this conceptual limitation, some alternative \emph{elastic measures} have been proposed to allow for time warping and provide a more flexible comparison. 

One of the most famous concepts for defining elastic measures is Dynamic Time Warping (DTW) \cite{berndt94}. This approach uses a warping path to align two series before their comparison, which is a sequence of adjacency matrix indices that defines a mapping, and the optimal path is the one that minimizes the global warping cost. The Complexity Invariant Distance (CID) \cite{batista11} is another shape-based measure that involves the Euclidean distance corrected by a complexity estimate of the series. The DISSIM \cite{frentzos07} and Short Time Series (STS) \cite{moller03} distances can deal with time series with different sampling rates by considering linear approximations of the series. DISSIM measures the difference by calculating the integral of the Euclidean distance between them while the STS measure considers the difference of the slopes between measurement points.

\paragraph{Edit-based distance functions.}

The idea behind these distance functions is commonly used to compare strings (Levenshtein distance) and consists of counting the minimum number of character insertions, deletions or substitutions to transform one string into another. The Edit Distance for Real Sequences (EDR) \cite{chen05} considers a threshold value that defines if two real values match or not. The difference is measured by the minimum number of operations to transform one series into another. EDR permits gaps between the two matched subsequences, but assigns penalties according to the lengths of the gaps. The Longest Common Subsequence (LCSS) distance \cite{vlachos02} uses an approach similar to the EDR. This measure permits time warping, as DTW, but also allows for gaps in comparison. Different from EDR, the LCSS does not assign penalties to the gaps.

\paragraph{Feature-based distance functions.}

These measures extract descriptive features from the time series and compare them. The Pearson correlation coefficient (PCC) $\rho$ is probably the most famous measure of this type. The PCC is defined as the covariance of the two time series divided by the product of their standard deviations. The Temporal Correlation and Raw Values (CORT) measure \cite{chouakria07} modulates a conventional distance (like Euclidean or DTW) according to the temporal correlation. Mutual information (MI) is another well known nonlinear measure defined as the information that two time series share \cite{meila03}. The maximal information coefficient (MIC) considers that if there exits a relationship between two time series, then it is possible to draw a grid on the scatter plot of these two series that encapsulates that relationship \cite{reshef11}. MIC automatically finds the best grid and returns the largest possible MI. Other examples of feature-based measures are based on cross-correlation functions \cite{deza09}, Fourier coefficients \cite{agrawal93}, auto-correlation coefficients \cite{galeano00} and periodograms \cite{casado03,caiado06}. 

\paragraph{Structure-based distance functions.}

Different from feature-based measures, these functions identify and compare high-level structures in the series. Most of them use parametric models to represent time series, e.g., auto-regressive moving average (ARMA) or Hidden Markov Models (HMM). Compression-based measures (CDM) \cite{cilibrasi05,keogh07} can also be considered to belong to the structure-based distance functions. These measures use compressors like gzip or bzip2 to compress a time series. The distance between two time series is proportional to the difference between the sizes of the compressed files.

\subsection{Network generation and analysis}
\label{subsec:methods}

We use the National Center for Environmental Prediction and National Center for Atmospheric Research (NCEP/NCAR) reanalysis dataset \cite{kalnay96} of monthly mean near-surface air temperature records on a global regular grid with $2.5\degree\times 2.5\degree$ resolution in latitude and longitude, spanning the period from 1948 to 2016. The resulting data set is composed of 10,512 time series with 816 values each. Prior to our analysis, we remove the seasonal component in each time series by using an additive decomposition by moving averages \cite{kendall83}. We consider here an additive model because the seasonal fluctuations are relatively constant over time. Without removing seasonality, the time series distance functions would show rather trivial patterns reflecting the dominant annual cycle component, while the focus of our present work is on finding similarities in other time series components such as  trends or low-frequency variability. For mathematical convenience, we finally normalize all time series to values between 0 and 1.

After preprocessing the data, we calculate the distance between every pair of time series. This step results in a distance matrix $\mathbf{D}$ whose element $D_{ij}$ contains the distance between the series $X_i$ and $X_j$. For this process, we consider each of the 29 distance functions listed in Tab.~\ref{tab:measures}. The definitions of all distance functions used in this paper can be found in Appendix~\ref{appendix:ts_ds} in Tab.~\ref{tab:measures_definitions_1}, \ref{tab:measures_definitions_2} and \ref{tab:measures_definitions_3}.

\begin{table}[!ht]
  \centering
  \footnotesize
  \caption{Time series distance functions used in the experiments} 
  \label{tab:measures}
  \begin{tabular}{llcc}
    \toprule
    & Distance & PDF & Reference \\
    \midrule        
    \rowcolor{gray!7}
    01 & Autocorrelation Coefficients (ACF) & & \cite{galeano00} \\
    02 & Avg($L_1$, $L_\infty$) (avgL1LInf) & \xmark & \cite{cha07} \\
    \rowcolor{gray!7}
    03 & Bhattacharyya & \xmark & \cite{bhattacharyya46} \\
    04 & Compression-based (CDM) & & \cite{keogh07} \\
    \rowcolor{gray!7}
    05 & Complexity Invariant (CID) & & \cite{batista11} \\
    06 & Correlation & & \cite{deza09} \\
    \rowcolor{gray!7}
    07 & Cross-correlation & & \cite{deza09} \\
    08 & Dice & \xmark & \cite{dice45} \\
    \rowcolor{gray!7}
    09 & DISSIM & & \cite{frentzos07} \\
    10 & Dynamic Time Warping (DTW) & & \cite{berndt94} \\
    \rowcolor{gray!7}
    11 & Euclidean ($L_2$) & & \cite{deza09} \\
    12 & Fourier Coefficient (fourierDist) & & \cite{agrawal93} \\
    \rowcolor{gray!7}
    13 & Gower & \xmark & \cite{gower71} \\
    14 & Integrated Periodogram (INTPER) & & \cite{casado03} \\
    \rowcolor{gray!7}
    15 & Jaccard & \xmark & \cite{deza09} \\
    16 & Kulczynski & \xmark & \cite{deza09} \\
    \rowcolor{gray!7}
    17 & Lorentzian & \xmark & \cite{deza09}\\
    18 & Manhattan & & \cite{deza09} \\
    \rowcolor{gray!7}
    19 & Mutual Information (MI) & \xmark & \cite{meila03} \\
    20 & Maximal Information Coefficient (MIC) & \xmark & \cite{reshef11} \\
    \rowcolor{gray!7}
    21 & Normalized Compression (NCD) & & \cite{cilibrasi05} \\ 
    22 & Partial Autocorrelation Coefficients (PACF) & & \cite{galeano00} \\
    \rowcolor{gray!7}
    23 & Periodogram (PER) & & \cite{caiado06} \\
    24 & Sorensen & \xmark & \cite{sorensen48} \\
    \rowcolor{gray!7}
    25 & Squared-Euclidean (sqdEuclidean) & \xmark & \cite{deza09} \\
    26 & Short Time Series (STS) & & \cite{moller03} \\
    \rowcolor{gray!7}
    27 & Tanimoto & \xmark & \cite{tanimoto58} \\
    28 & Temporal Correlation and Raw Values (CORT) & & \cite{chouakria07} \\
    \rowcolor{gray!7}
    29 & Wave Hedges & \xmark & \cite{cha07} \\
    \bottomrule
  \end{tabular}
\end{table}

For the sake of coherence, we use exclusively distance functions in this paper instead of explicitly considering similarity measures as well. Notably, some functions, like the Pearson correlation coefficient ($\rho$) and mutual information (MI), are actually similarity functions that return high values when two time series are similar. Instead of using $\rho$ directly, we define an associated distance form $d_{\text{COR}}(X,Y) = 1 - |\rho_{XY}|$. Following the same approach as in previous works \cite{donges09}, we use here the absolute value of the Pearson correlation coefficient because both large negative and large positive correlation values indicate strong statistical interdependence. We apply the same strategy for other similarity measures defined in the range $[0,1]$. The cross-correlation distance is calculated by taking the largest absolute correlation distance for lags in the interval $[-24, +24]$ months. We considered a maximum delay of two years to capture associations delayed by trimesters, semesters, or a year. Conversely, we did not consider longer lags to avoid the drastic reduction in the number of values in the lagged time series. The mutual information (MI) is a similarity function defined in the range $[0,+\infty)$ but commonly bound from above by a maximum value that is specific to the particular estimator used. Instead of using MI itself, we therefore employ here an MI-based distance measure \cite{meila03} with values in the unit interval $[0,1]$. 

Some distance functions measure the dissimilarity by comparing the probability density functions (PDF) of two time series (measures with the PDF column checked in Tab.~\ref{tab:measures}). For these measures, we follow the approach presented in \cite{cha07} that consists of discretizing every normalized series into 64 bins and dividing the frequencies by the length of the series. The ranges for all the bins are exactly the same because all series were previously normalized to unit range. 

After calculating the distance matrix $\mathbf{D}$ for every distance function, we normalize it to simplify the comparison between different measures. Then, we use $\mathbf{D}$ to construct a binary adjacency matrix $\mathbf{A}$ by applying a threshold value $\tau$ as
\begin{equation}
    \label{eq:dist_matrix_threshold}
    \mathbf{A} = \Theta(\tau - \mathbf{D}),
\end{equation}
\noindent 
where $\Theta(\bullet)$ is the Heaviside step function. Choosing a proper threshold value $\tau$ is fundamental for the network construction process to provide meaningful results. Higher values ($\tau \to 1$) will generate densely connected networks while lower ones ($\tau \to 0$) create fragmented networks. For each distance function, we calculate the distance matrix $\mathbf{D}$ and use the $p^{th}$ distance percentile as the threshold value ($\tau$) to build the network. The distance percentile $p$ is a single parameter that controls the edge density of all networks. Advantages of this approach include the unneccessity to choose a distinct $\tau$ for each distance function and the creation of networks with the same number of edges. Thus, we avoid the topological comparison of networks with different numbers of edges. 

As a result of the described process, we obtain a single undirected and unweighted network for every distance function, all of them exhibiting the same number of edges. We compare these networks by using the Hamming distance  \cite{levenshtein66}, which measures the number of edges that have to be added or deleted to transform one network into another. Given two labelled networks $G$ and $G'$, where $A^{G}$ and $A^{G'}$ are their respective adjacency matrices, the Hamming distance is defined as
\begin{equation}
    \label{eq:hamming}
    d_H(A^{G}, A^{G'}) = \sum_{i,j}\frac{1}{2}\mathrm{xor}(A^{G}_{i,j},A^{G'}_{i,j}),    
\end{equation}
\noindent 
where $\mathrm{xor}(\bullet,\bullet)$ is the logical exclusive-or operator. 

For defining spatial distances between grid points on a spherical grid, we use the Haversine distance
\begin{equation}
    \label{eq:haversine}
    d_{Haversine} ((\phi_1,\lambda_1), (\phi_2,\lambda_2)) =  2r\arcsin \Bigg(\sqrt{\sin^2 \bigg(\frac{\phi_2-\phi_1}{2}\bigg) + \cos(\phi_1)\cos(\phi_2)\sin^2\bigg(\frac{\lambda_2-\lambda_1}{2}\bigg)}\Bigg),
\end{equation}
\noindent
where $\phi$ and $\lambda$ correspond to the latitude and longitude of the two locations respectively, and $r$ is the Earth's radius ($r\approx 6378137$~m).

As a particular scalar characteristic of the obtained networks, we consider the smallworldness index \cite{humphries08} defined as
\begin{equation}
    \label{eq:smallworldness}
    SWI=\frac{C(G)/C(G_{rand})}{L(G)/L(G_{rand})},
\end{equation}
\noindent 
where $C(\bullet)$ and $L(\bullet)$ are the respective global clustering coefficient and average shortest path length of the graph $G$ and a correspondent random graph $G_{rand}$, respectively. Considering that this measure uses a probabilistic procedure to generate the random graph benchmark, we repeat this process 50 times and report the mean value taken over all realizations. 

One of the concepts used in this paper is a teleconnection, which we define here as the most similar pairs of time series whose grid cells have great-circle distances longer than $5000$~km \cite{glantz91}. By contrast, we disregard short-distance connections since they tend to be similar because of the spatial continuity of the studied temperature field commonly resulting in small time series distances between close nodes. In our analysis, we use the Haversine distance to compute the great-circle distance between two coordinates (Eq.~\ref{eq:haversine}). Specifically, we use the center coordinates of each grid cell to define this distance. For keeping the number of results interpretable, we also limit the number of distance functions considered. Instead of studying the networks generated by all the 29 distance functions, we limit ourselves to eight since many distance functions generate similar networks. We have chosen these distance functions by applying hierarchical clustering (Ward's method) to the Hamming distances between the adjacency matrices of all combinations of networks. After obtaining the eight clusters, we select the centroid network exhibiting the shortest average distance to the other networks in a given cluster. 

The choice of $p$ does not change substantially the teleconnections since this is a lower-bound threshold by definition. It means that, for $p=0.01$ or higher values, the strongest links are the same and they all lead to the same teleconnections. Since we show the 500 strongest links (shortest distances), $p$ would just have an impact if it was so small that the network would had less than 500 links. But in this case, $p$ is so small that the network is very disconnected and does not properly represent the system. Therefore, the choice of $p$ is not critical for determining the teleconnections but it has an impact on the networks clustering and on the cluster network centroids. In this paper, we focus on the network centroids as major representatives of the time series distance functions in the cluster but we include the teleconnections for all networks in the Appendix~\ref{append:teleconnections} for the interested reader.

\section{Results and discussion}
\label{sec:results}

In the following, we evaluate the topological differences of the networks constructed using different time series distance functions. For a review on complex network theory and the full definitions of all measures used throughout this paper, we refer the interested reader to existing review papers of this subject \cite{barabasi16,boccaletti06,costa07}. 

\subsection{Edge density effects}

\begin{figure}[!ht]
\centering
\includegraphics[width=\linewidth]{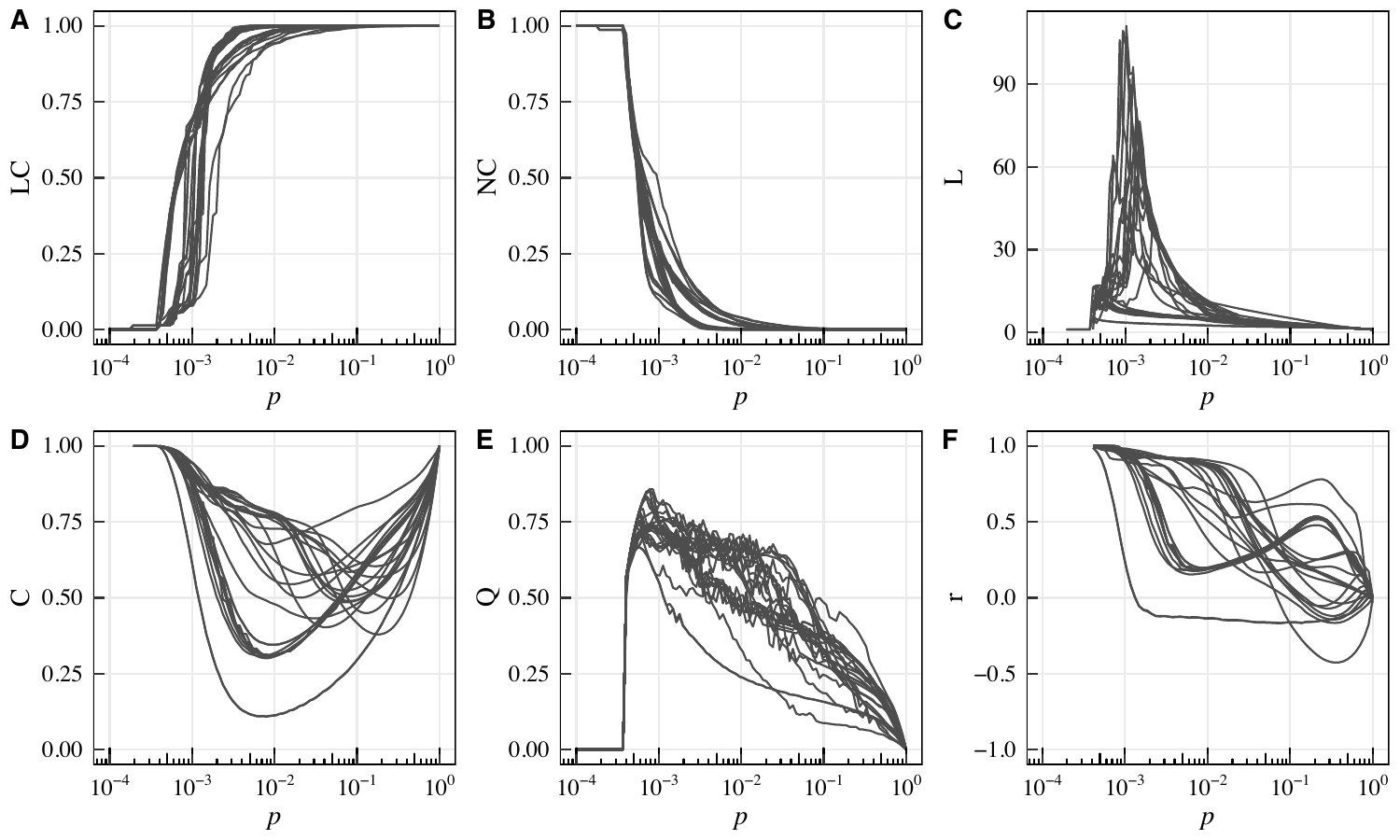}
\caption{Influence of the edge density $p$ on global network measures: (A) normalized size of the largest component, (B) normalized number of components in the network, (C) average path length $L$, (D) clustering coefficient $C$, (E) modularity $Q$, (F) assortativity $r$. The different lines correspond to the 29 different distance functions used in this work.}
\label{fig:net_global_measures}
\end{figure}

In our experiments, we generate and compare networks with a given edge density $p$. To obtain a better understanding of the effect of this parameter, we first study the corresponding behavior of some important global network measures for different distance functions, which will provide the basis for selecting appropriate values of $p$ for the following  analyses. Our results are summarized in Fig.~\ref{fig:net_global_measures}. 

When the edge density is very low ($p \lesssim 0.0004$), only a few extremely similar pairs of time series are connected, resulting in a very small number of edges and many disconnected components. In technical terms, we have not yet reached the percolation threshold of the network, which can be seen in terms of the very small size of the largest component along with the large number of network components. In this regime, we have a low average path length $L$ (distance between connected nodes), since the majority of nodes belonging to the same component are in fact directly connected. For the same reason, we also find a high global clustering coefficient $C$ (density of closed triangles in the network) and low network modularity $Q$ (heterogeneity associated with community structures). At $p\sim\mathcal{O}(10^{-3})$, we observe a marked change of the network connectivity resulting in the emergence of a giant component along with a marked reduction in the overall number of network components, which is associated with the percolation transition. Notably, this transition occurs at an edge density that is about one order of magnitude larger than for an Erd\"os-R\'enyi random graph of the same size ($p_c\sim 1/N$), pointing to the fact that edges are not generated at random, but respect the spatial embedding of the network and possibly also tend to cluster in space.

Around the percolation transition, we observe a sharp peak of $L$ and a marked decrease of $C$ at $p\sim\mathcal{O}(10^{-3})$. Both effects are directly related with the increasing connectivity of the network along with the emergence of a giant component. Notably, more and more nodes become (indirectly) connected in the network and thereby contribute to the calculation of both measures. During this phase (about $0.0004\lesssim p\lesssim 0.1$, nodes tend to connect with others of a similar degree (positive assortativity), except for the network created using the CDM distance function that is weakly disassortative (lowest line in Fig.~\ref{fig:net_global_measures}F). When $p \approx 0.01$, the networks are almost connected (with only very few isolated nodes) but still sparse. When $p \gtrsim 0.01$, more and more closed triangles appear, and $C$ increases towards its theoretical maximum value ($C=1$) obtained for a fully connected network ($p=1$). On the other hand, the modularity decreases when the network becomes dense since there are no well defined communities anymore. When $p \gtrsim 0.01$, the networks show apparent small-world features (high $C$ and low $L$), which however can be explained by the spatial embedding of the graph as emphasized in several previous studies \cite{Bialonski2010,Bialonski2011,Hlinka2012}. 

In summary, the interval $0.0004 \lesssim p \lesssim 0.1$ corresponds to a transition from a disconnected to a strongly connected climate network. After passing this transition phase, the number of edges can further increase until the networks become fully connected ($p=1$). Networks at high edge densities are however not practically interesting since they connect almost all pairs of time series, including those that are not similar, and generate alike networks. Therefore, we consider that $0.0004 \lesssim p \lesssim 0.1$ is an interesting range of edge densities because only the most similar pairs of time series are connected and the networks are not fragmented into too many components. This is also well in line with the settings used in the vast majority of previous climate network studies. For a more in-depth analysis, we have selected four values of $p$ based on its influence on the network topology: $0.001, 0.01, 0.05$, and $0.1$. The results of our following analyses will be reported for those four cases.

\subsection{Global network features}

\begin{figure}[!htb]
  \centering
  \includegraphics[width=0.94\linewidth]{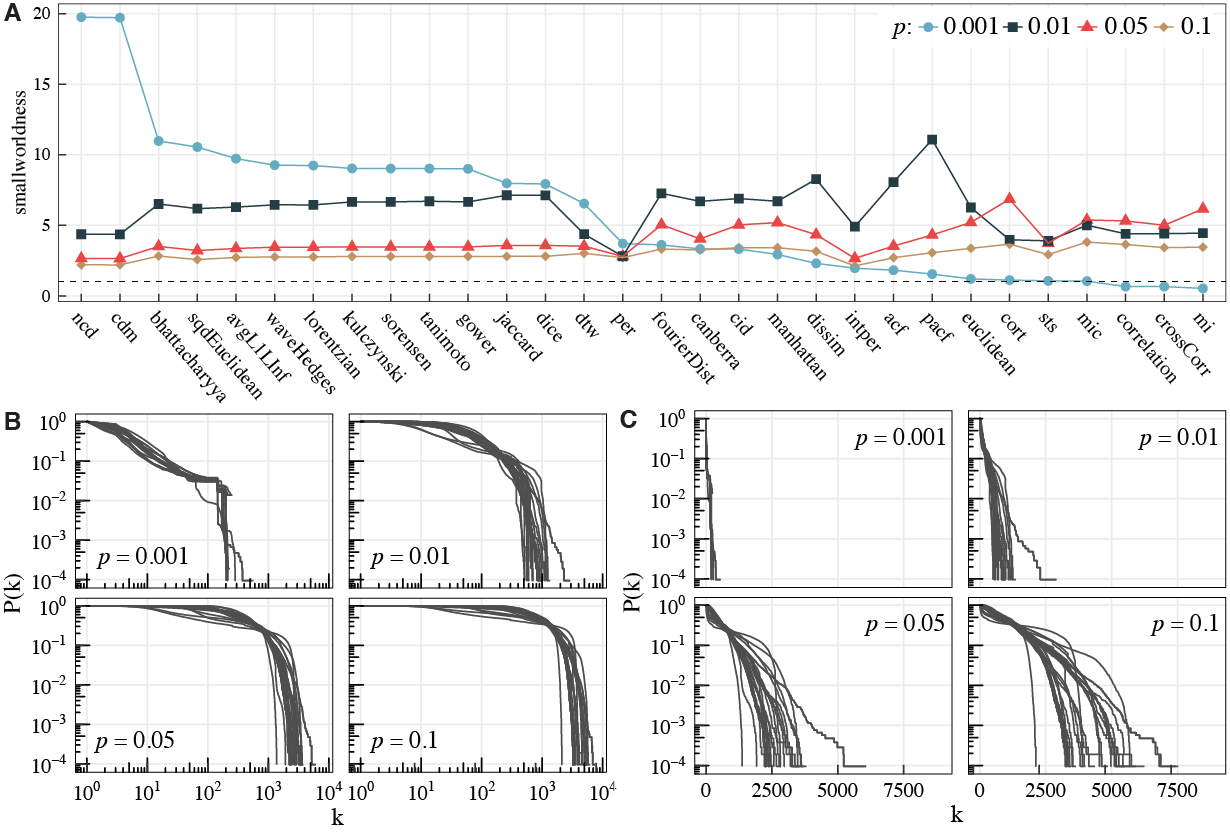}
  \caption{(A) Smallworldness index SWI (Eq.~\ref{eq:smallworldness}) for the networks created using different distance functions (see Tab.~\ref{tab:measures}) and four different edge densities. Values higher than one (dashed line) indicate that the network exhibits the small-world effect. Our results show that almost all networks present the small-world feature, except for $p=0.001$, in which case some networks rather resemble random graphs (SWI$\approx$ 0). Panels (B) and (C) show the cumulative degree distributions in log-log scale and linear-log scale, respectively. The results suggest that, for small edge densities ($p=0.001$), the degree distributions follow a power law with sharp cutoff, while higher values of $p$ result in degree distributions decaying faster than a power law.}
  \label{fig:network_models} 
\end{figure}

In the following, we further analyze a few selected network properties that are commonly used to classify complex network into certain categories like scale-free and small-world networks \cite{barabasi16}.  Figure~\ref{fig:network_models}A shows the values of the smallworldness index \cite{humphries08} for the different distance functions and four values of the edge density. According to its definition, values of the smallworldness index larger than one (dashed line) are indicative of the small-world feature (i.e., high $C$ along with low $L$). The exact values for the network measures can be found in Appendix~\ref{appendix:climate_nets_measures}, Tab.~\ref{tab:climate_net_measures_1}, \ref{tab:climate_net_measures_2}, \ref{tab:climate_net_measures_3} and \ref{tab:climate_net_measures_4}. Our results show that all generated networks indeed exhibit the small-world property, except for the correlation, cross-correlation, mutual information, MIC, STS and CORT functions when $p= 0.001$. These results are in agreement with 
previous studies reporting a small-world effect in climate networks \cite{donges09,tsonis04}. Our results show that this effect is also present when other distance functions are considered. On the one hand, the high clustering coefficient results from the spatial continuity of the underlying physical fields \cite{donges09} which tends to connect close cells (nodes) and thereby form triangles. On the other hand, the existence of long-range connections (teleconnections) explains the small average path length. These long distance edges are responsible for an efficient propagation of local perturbations through the whole network, as will be further discussed in Section~\ref{subsec:results_teleconnections}. We emphasize again that one reason for the small-world effect to emerge across all climate networks is the spatial embedding of the nodes. Although we found evidences of small-worldness for almost all distance functions, some or all of them might be biased, as demonstrated by \cite{hlinka17}. Studying whether or not this already explains the observed characteristics completely would be possible by using sophisticated spatial network surrogates \cite{Wiedermann2016}, the consideration of which is however beyond the scope of the present work.

Small-world networks like those presented by the studied climate networks can be further classified according to their degree distributions into three types: scale-free, broad-scale and single-scale \cite{amaral20}. Scale-free networks have a power law distribution, broad-scale networks have a power law regime followed by a sharp cutoff, and single-scale networks present a faster than algebraic decay, e.g., exponential or Gaussian. In Fig.~\ref{fig:network_models}B and C, we show the cumulative degree distributions for all networks separated by edge densities in a log-log and linear-log scales, respectively. Our results indicate that all distance functions generate broad-scale networks for small edges densities ($p=0.001$), i.e., they present a scale-free interval followed by sharp decays. For larger edge densities, all distance functions generate single-scale networks.

\subsection{Network similarities}
\label{subsec:net_similarity} 

We next investigate the similarities in the network structures created using different time series distance functions. As in our previous analyses, we consider four edge densities: $0.001, 0.01, 0.05$, and $0.1$. For each value of $p$, we construct the networks using all 29 distance functions (Tab~\ref{tab:measures}) and compare the resulting networks using the Hamming distance \cite{levenshtein66} (Eq.~\ref{eq:hamming}). Therefrom, we build a distance matrix $\mathbf{D}^{H}$ whose element $D^{H}_{A,B}$ is the Hamming distance $d_H(A,B)$ between the networks $A$ and $B$. Then, we use hierarchical agglomerative clustering (using Ward's method) to cluster the different networks. 

\begin{figure}
  \centering
  \includegraphics[width=0.85\linewidth]{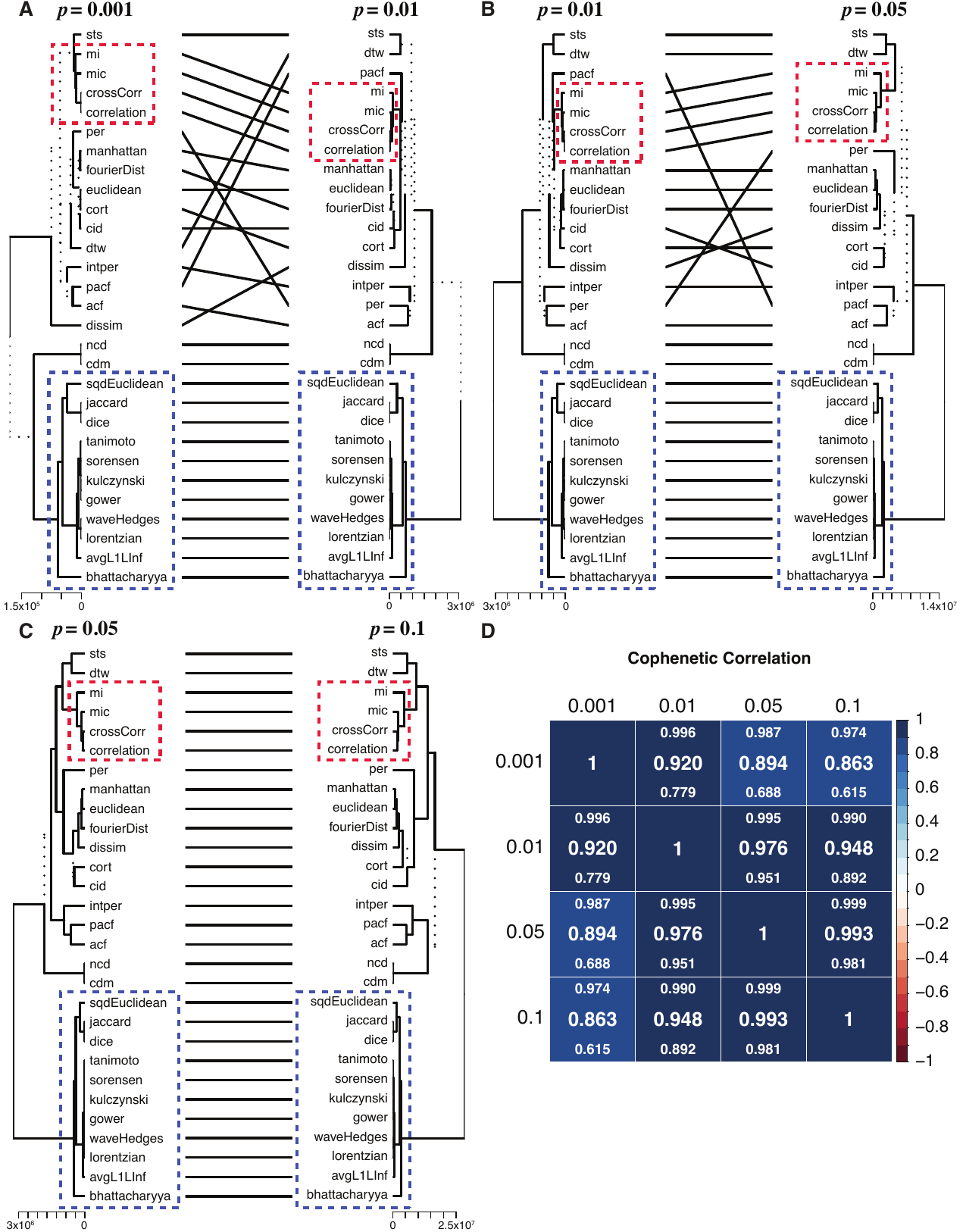}
  \caption{(A)-(C) Dendrograms created by applying hierarchical clustering (Ward's method) to the Hamming distances between networks created using different time series distance functions and four edge densities ($0.001, 0.01, 0.05$, and $0.1$). The lines connecting labels from two dendrograms show the different positions of the same label. Strong entanglement between the lines indicates low similarity between the dendrograms. The dashed lines highlight the difference between two dendrograms. (D) Cophenetic correlation measuring the similarity between the dendrograms. Values close to 1 indicate high similarity while values close to 0 means no similarity. The upper and lower values inside each cell are the upper and lower limit values from a 99\% bootstrap confidence interval. For this test, we randomly permuted the labels of the dendrogram without modifying the topology. We repeated this process 1000 times to build the confidence levels. These results show that the clustering results for different edge densities are strongly correlated, especially for higher values of $p$.}
  \label{fig:dendrograms_comparison}
\end{figure}

The obtained clustering results are shown in Fig.~\ref{fig:dendrograms_comparison}. The distance matrices for the four edge densities can be found in Appendix~\ref{appendix:dnet_distances} (Fig.~\ref{fig:net_dists}). Our results suggest that, for all four edge densities, all PDF-based distance functions (Tab.~\ref{tab:measures}) except the mutual information form one cluster (blue dotted lines in Fig.~\ref{fig:dendrograms_comparison}). This indicates that all of them generate similar networks independent of $p$. In the literature, a large number of studies has constructed functional climate networks using Pearson correlation with or without lag, while a minority has used mutual information (MI) \cite{donges09,jingfang17,tsonis04,guez14,tsonis08}. According to our results, these three approaches generate similar networks (red dotted lines in Fig.~\ref{fig:dendrograms_comparison}), which confirms previous results \cite{donges09,Radebach2013}. We also find that the Pearson correlation and cross-correlation generate similar networks, which suggests that previous studies using both measures are compatible and that the lag does not markedly affect the network topology.

Figure~\ref{fig:dendrograms_comparison} also presents a comparison between the dendrograms and their cophenetic correlations. The idea here is to verify if the clustering hierarchies are consistent for different edge densities. The black lines between dendrograms indicate the positions of one network in different dendrograms. Strong entanglement indicates large differences between the dendrograms. Our results show that there exists some entanglement between $p = 0.001$, $0.01$ and $0.05$ (Fig.~\ref{fig:dendrograms_comparison}A and B) that is mainly caused by the small number of links in the networks. Even though, the clustering hierarchies can be considered similar since their cophenetic correlations are high ($> 0.89$), especially at higher edge densities (Fig.~\ref{fig:dendrograms_comparison}C). 

\subsection{Teleconnections}
\label{subsec:results_teleconnections}

We have already shown that networks constructed with different distance functions exhibit distinct topologies. We have also observed that almost all resulting networks display the small-world feature. If the networks are small-world and have different topology, the long-distance edges responsible for the small-world feature should be different. This indicates that distinct distance functions can capture different long-distance relationships which can be caused by different underlying climate effect.

In the following, we further analyze the differences between the topologies generated by different distance functions. As described in Sec.~\ref{subsec:methods}, we define teleconnections as the most similar pairs of time series whose grid cells have great-circle distances longer than $5000$~km \cite{glantz91}. Atmospheric teleconnection patterns can be generated by both, internal atmospheric processes and forcing from surface conditions. According to \cite{liu2007atmospheric}, such teleconnections enable the atmosphere to act like a "bridge" between different parts of the ocean and while enabling the ocean to act like a "tunnel" linking different atmospheric regions. These long-distance links are of great importance for climate science because they can provide significant skill to statistical weather forecasts from synoptic to seasonal scales and even beyond \cite{alexander02,boers19}.

Our methodology (Sec.~\ref{subsec:methods}) consisted on clustering the network topologies and extracting the centroid network of each of cluster. Figure~\ref{fig:teleconnections} shows the 500 strongest teleconnection related links for the eight different centroid networks. As already demonstrated in previous works \cite{donges09,tsonis08,Radebach2013}, the correlation captures especially the influence of the El Niño-Southern Oscillation (ENSO). ENSO is a very important teleconnection pattern responsible for climate variability all over the globe \cite{palus11,nigam15,dong15}. This phenomenon originates from nonlinear ocean-atmosphere interactions in the tropical Pacific ocean connecting large-scale oceanic sea surface temperature (SST) anomalies (El Niño) with the large-scale atmospheric Southern Oscillation (SO), which is characterized by a strong sea-level pressure seesaw between French Polynesia and North Australia \cite{picaut97,gong1999definition}. This pattern is also captured by Euclidean, DTW, INTPER, and ACF distances. The teleconnections found by CDM and Gower distances do not exhibit coherent spatial patterns, which is expected in physical mechanisms \cite{boers19}. This lack of coherence does not imply that patterns do not exist but that we cannot identify any using our methodology. 


\afterpage{%
  \begin{figure}[h!]
    \centering
    \includegraphics[width=\linewidth]{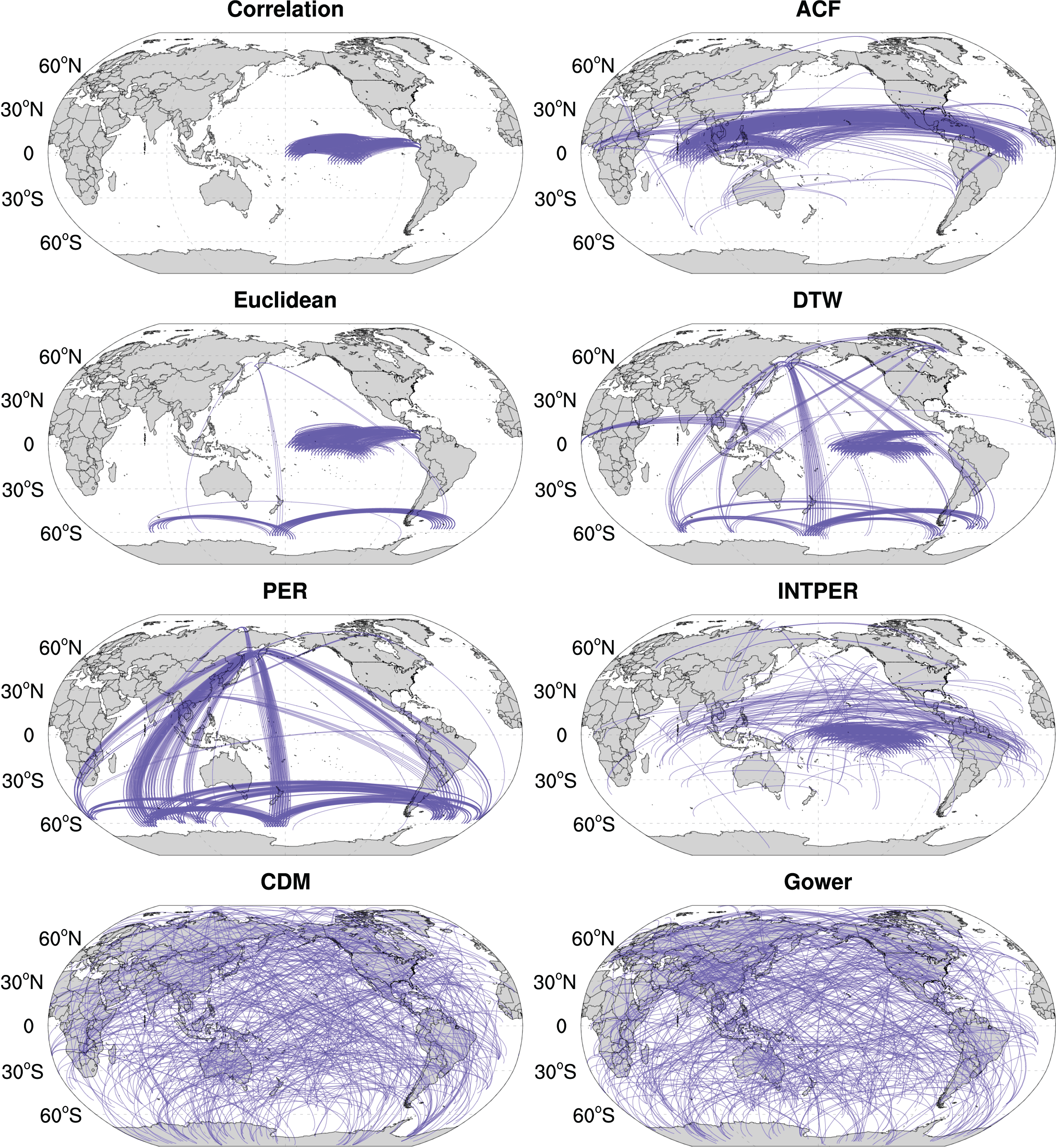}
    \caption{Teleconnections for the eight distance functions resulted from the cluster analysis: Correlation, ACF, Euclidean, DTW, PER, INTPER, CDM, and Gower. We considered only the 500 strongest teleconnection links (nodes pairs with distance greater than $5000$~km).}
    \label{fig:teleconnections} 
  \end{figure}
  \clearpage
}

The Euclidean, DTW, and PER distances also capture teleconnections similar to those associated with the Pacific-South America (PSA) pattern and the Antarctic Oscillation (AAO). The PSA modes are unique features of atmospheric variability in the Southern Hemisphere in the subpolar/polar regions of the South Pacific. The PSA pattern was suggested as part of a stationary Rossby wave train, which is usually generated by changes in tropical convection \cite{mo1998pacific}. According to \cite{mo01}, the PSA pattern is related to sea surface temperature (SST) variability over the central and eastern Pacific at decadal scales, and it is a response to the El Niño–-Southern Oscillation (ENSO) at interannual time scales. The PSA can also be associated with the quasi-biennial component of ENSO, and the strongest connections occur during austral spring. This pattern represents a zonal wave with wavenumber 3 and a well-defined wave train from the tropical Pacific and the Indian Ocean to South America with a large amplitude in the PSA sector \cite{mo01}. The zonal wavenumber represents the number of waves at a given latitude. The large amplitude in the PSA sector means that the cold air advance over a region, with impacts in the regional climate \cite{renwick99}. PSA is also related to the AAO, which is characterized by zonally symmetric structures with opposite signs between the Antarctic region and latitudes of about 45$\degree$S \cite{thompson2000annular}. The annular pattern of AAO can be considered as a presence of anomalies with latitudinal variation between the regions of extratropical and subtropical latitudes of the Southern Hemisphere \cite{kidson1988indices}. The positive phase occurs when there is a predominance of negative SST anomalies close to Antarctica, and positive in the mid-latitudes, resulting in a shift of extratropical cyclones and cold fronts towards Antarctica \cite{vasconcellos2010extreme}.

Through INTPER, it is possible to observe the relationship between the eastern Pacific and tropical Atlantic temperatures that can be explained by the Southern Oscillation \cite{aceituno88}. The ACF and INTPER show an interesting inter-basin connection between the Indian, Pacific and Atlantic oceans. This connection can be explained by the Indian Ocean Dipole (IOD). The positive phase of the IOD is associated with warmer (colder) than normal SST off the coast of East Africa (Java-Sumatra) \cite{chan2008indian}. The signal of the associated atmospheric teleconnection can be noticed far from the source region \cite{saji2003possible,yamagata2004coupled,behera2005paramount}, also affecting the SSTs in the Atlantic Ocean near South America \cite{chan2008indian}. Empirical orthogonal function (EOF) analysis indicates that 30\% of the total variation of anomalous Indian Ocean SSTs is during ENSO events and the dipole mode (IOD) explains about 12\% of the associated variability \cite{saji1999dipole}. 


Our results further show that different distance functions are capable to capture different patterns and relationships that are not very well understood, such as those shown in ACF and INTPER based networks. These patterns represent an enhancement of existing knowledge and should be further investigated. Furthermore, some distance functions also indicate interesting interactions in the meridional direction, connecting the two hemispheres. Those interactions are not trivial as the atmosphere and ocean tend to flow zonally as constrained by the conservation of angular momentum \cite{liu2007atmospheric}, which suggest that studies to explore these connections can be made through the presented functions.

\section{Conclusions}
\label{sec:conclusion}

In this work, we have analyzed the effect of 29 different time series distance functions on the topology of functional climate networks. We have started by studying how the edge density $p$ affects the resulting network topology. When $p \lesssim 0.0004$, only a few very similar nodes are connected and the networks present many components. On the other hand, when $p \gtrsim 0.01$ the networks become densely connected and even not very similar nodes are mutually linked. The interval $0.0004 \lesssim p \lesssim 0.1$ manifests a transition phase where the networks get more and more connected. We consider this interval as relevant for the study of climate networks, at least for the specific dataset employed in our work. The resulting networks exhibit a (likely geometrically induced) small-world effect and have a degree distribution that decays faster than a power law distribution. 

A large body of previous studies have used the Pearson correlation coefficient, covariance or cross-correlation to measure similarity between time series. We have shown here that these measures generate similar networks. Our results also demonstrate that other distance functions are capable of generating different network topologies. As all these networks exhibit similar small-world characteristics, yet with distinct topologies, the long-range connections responsible for this effect tend to be different.

It is important to remember that, although we found evidences of small-worldness for almost  all distance functions, this effect may be biased by the network construction process \cite{hlinka17}, which represents a limitation in the methodology here proposed. Nevertheless, we have observed that distinct distance functions can capture different teleconnection patterns. These teleconnections play an import role in the climate system since they generally represent energy transfer between remote regions \cite{dong15} and can be used for improving statistical weather forecasts at various time scales \cite{alexander02,boers19}. Although the structure of several teleconnections patterns has been known, the reasons why they emerge are not always well understood \cite{nigam15}. 

One advantage of functional climate networks is the possibility to inspect the climate dynamics from a complementary perspective. We have shown here that alternative distance functions capture different teleconnections that present patterns similar to some known climate phenomena like the Pacific-South America (PSA) pattern, Antarctic Oscillation (AAO), and Indian Ocean Dipole (IOD). Phenomena like these have not been investigated in previous works using climate networks. A few distance functions, such as CDM and Gower (Fig. \ref{fig:teleconnections}), did not present coherent spatial patterns. It does not mean that a pattern does not exist, but that we cannot identify any using our methodology. We believe that this is not a limitation in our methodology, but an indicative that some distance functions can be inappropriate for the study of climate. To conclude, we believe that our results provide a new perspective for climate science because they allow the investigation of climate patterns from a different point of view. This advanced approach could finally lead to a better understanding of how global climate patterns emerge.


In summary, we have observed differences in the topologies of climate networks generated by different distance functions. These distance functions capture non-trivial climatological phenomena that previous studies on climate network have not found. This research opens many opportunities for future works. One obvious next step is the thorough evaluation of the climate patterns and the deep reasons behind the topological differences. This  analysis can be easily extended to different climate variables using the distance functions used in this study. Other distance functions may also be considered, for example, synchronization-based measures and causality analysis can be applied. Ultimately, taking temporal variations in the corresponding climate patterns into account, such an investigation can also be extended into a temporal network framework.  

\section*{Acknowledgments}

This research received financial support from the S\~ao Paulo Research Foundation (FAPESP) grants 2017/05831-9 and 2015/50122-0. The authors also acknowledge the National Council for Scientific and Technological Development (CNPq) for its financial support. This research was developed using computational resources from the Center for Mathematical Sciences Applied to Industry (CeMEAI) funded by FAPESP (grant 2013/07375-0). RVD has received funding by the German Federal Ministry for Education and Research (BMBF) via the BMBF Young Investigators Group CoSy-CC$^2$ (Complex Systems Approaches to Understanding Causes and Consequences of Past, Present and Future Climate Change, grant no.~01LN1306A), the Belmont Forum / JPI Climate project GOTHAM (Globally Observed Teleconnections and Their Representation in Hierarchies of Atmospheric Models, grant no.~01LP16MA) and the JPI Climate / JPI Oceans project ROADMAP (The Role of Ocean Dynamics and Ocean-Atmosphere Interactions in Driving ClimAte Variations and Future Projections of Impact-Relevant Extreme Events, grant no. 01LP2002B).

\section*{Computer Code Availability}

The method and the experiments in this paper were implemented using the R programming language \cite{rstats} with the packages \texttt{igraph} \cite{igraph06}, \texttt{TSclust} \cite{tsclust14}, and \texttt{TSdist} \cite{mori16}. The R software, the packages, and our implementation are all open-source (GPL) and freely available for download. Our source codes can be obtained at \url{https://lnferreira.github.io/climate\_networks\_R}.



\bibliographystyle{model1-num-names}







\bibliography{ref.bib}

\begin{thebibliography}{82}
\expandafter\ifx\csname natexlab\endcsname\relax\def\natexlab#1{#1}\fi
\providecommand{\bibinfo}[2]{#2}
\ifx\xfnm\relax \def\xfnm[#1]{\unskip,\space#1}\fi
\bibitem[{Dijkstra et~al.(2019)Dijkstra, Hern\'andez-Garc\'ia, Masoller, and
  Barreiro}]{Dijkstra2019}
\bibinfo{author}{H.~A. Dijkstra}, \bibinfo{author}{E.~Hern\'andez-Garc\'ia},
  \bibinfo{author}{C.~Masoller}, \bibinfo{author}{M.~Barreiro},
  \bibinfo{title}{{Networks in Climate}}, \bibinfo{publisher}{Cambridge
  University Press}, \bibinfo{address}{Cambridge}, \bibinfo{year}{2019}.
\bibitem[{Donner et~al.(2017)Donner, Wiedermann, and Donges}]{Donner2017}
\bibinfo{author}{R.~V. Donner}, \bibinfo{author}{M.~Wiedermann},
  \bibinfo{author}{J.~F. Donges},
\newblock \bibinfo{title}{{Complex Network Techniques for Climatological Data
  Analysis}},
\newblock in: \bibinfo{editor}{C.~Franzke}, \bibinfo{editor}{T.~O'Kane} (Eds.),
  \bibinfo{booktitle}{Nonlinear and Stochastic Climate Dynamics},
  \bibinfo{publisher}{Cambridge University Press},
  \bibinfo{address}{Cambridge}, \bibinfo{edition}{1} edition,
  \bibinfo{year}{2017}, pp. \bibinfo{pages}{159--183}.
\bibitem[{Boers et~al.(2014)Boers, Bookhagen, Barbosa, Marwan, Kurths, and
  Marengo}]{boers14}
\bibinfo{author}{N.~Boers}, \bibinfo{author}{B.~Bookhagen},
  \bibinfo{author}{H.~M.~J. Barbosa}, \bibinfo{author}{N.~Marwan},
  \bibinfo{author}{J.~Kurths}, \bibinfo{author}{J.~A. Marengo},
\newblock \bibinfo{title}{Prediction of extreme floods in the eastern central
  andes based on a complex networks approach},
\newblock \bibinfo{journal}{Nature Communications} \bibinfo{volume}{5}
  (\bibinfo{year}{2014}) \bibinfo{pages}{5199}.
\bibitem[{Donges et~al.(2009)Donges, Zou, Marwan, and Kurths}]{donges09}
\bibinfo{author}{J.~F. Donges}, \bibinfo{author}{Y.~Zou},
  \bibinfo{author}{N.~Marwan}, \bibinfo{author}{J.~Kurths},
\newblock \bibinfo{title}{Complex networks in climate dynamics},
\newblock \bibinfo{journal}{The European Physical Journal Special Topics}
  \bibinfo{volume}{174} (\bibinfo{year}{2009}) \bibinfo{pages}{157--179}.
\bibitem[{Fan et~al.(2017)Fan, Meng, Ashkenazy, Havlin, and
  Schellnhuber}]{jingfang17}
\bibinfo{author}{J.~Fan}, \bibinfo{author}{J.~Meng},
  \bibinfo{author}{Y.~Ashkenazy}, \bibinfo{author}{S.~Havlin},
  \bibinfo{author}{H.~J. Schellnhuber},
\newblock \bibinfo{title}{Network analysis reveals strongly localized impacts
  of el ni{\~n}o},
\newblock \bibinfo{journal}{Proceedings of the National Academy of Sciences}
  \bibinfo{volume}{114} (\bibinfo{year}{2017}) \bibinfo{pages}{7543--7548}.
\bibitem[{Steinhaeuser et~al.(2011)Steinhaeuser, Chawla, and
  Ganguly}]{steinhaeuser11}
\bibinfo{author}{K.~Steinhaeuser}, \bibinfo{author}{N.~V. Chawla},
  \bibinfo{author}{A.~R. Ganguly},
\newblock \bibinfo{title}{Complex networks as a unified framework for
  descriptive analysis and predictive modeling in climate science},
\newblock \bibinfo{journal}{Statistical Analysis and Data Mining}
  \bibinfo{volume}{4} (\bibinfo{year}{2011}) \bibinfo{pages}{497--511}.
\bibitem[{Tsonis and Roebber(2004)}]{tsonis04}
\bibinfo{author}{A.~Tsonis}, \bibinfo{author}{P.~Roebber},
\newblock \bibinfo{title}{The architecture of the climate network},
\newblock \bibinfo{journal}{Physica A: Statistical Mechanics and its
  Applications} \bibinfo{volume}{333} (\bibinfo{year}{2004})
  \bibinfo{pages}{497 -- 504}.
\bibitem[{Ferreira et~al.(2020)Ferreira, Vega-Oliveros, Cotacallapa, Cardoso,
  Quiles, Zhao, and Macau}]{ferreira20}
\bibinfo{author}{L.~N. Ferreira}, \bibinfo{author}{D.~A. Vega-Oliveros},
  \bibinfo{author}{M.~Cotacallapa}, \bibinfo{author}{M.~F. Cardoso},
  \bibinfo{author}{M.~G. Quiles}, \bibinfo{author}{L.~Zhao},
  \bibinfo{author}{E.~E.~N. Macau},
\newblock \bibinfo{title}{Spatiotemporal data analysis with chronological
  networks},
\newblock \bibinfo{journal}{Nature Communications} \bibinfo{volume}{11}
  (\bibinfo{year}{2020}) \bibinfo{pages}{4036}.
\bibitem[{Guez et~al.(2014)Guez, Gozolchiani, and Havlin}]{guez14}
\bibinfo{author}{O.~C. Guez}, \bibinfo{author}{A.~Gozolchiani},
  \bibinfo{author}{S.~Havlin},
\newblock \bibinfo{title}{Influence of autocorrelation on the topology of the
  climate network},
\newblock \bibinfo{journal}{Phys. Rev. E} \bibinfo{volume}{90}
  (\bibinfo{year}{2014}) \bibinfo{pages}{062814}.
\bibitem[{Tsonis and Swanson(2008)}]{tsonis08}
\bibinfo{author}{A.~A. Tsonis}, \bibinfo{author}{K.~L. Swanson},
\newblock \bibinfo{title}{{Topology and Predictability of El Ni\~no and La
  Ni\~na Networks}},
\newblock \bibinfo{journal}{Physical Review Letters} \bibinfo{volume}{100}
  (\bibinfo{year}{2008}) \bibinfo{pages}{228502}.
\bibitem[{Palu{\v{s}}(2018)}]{palus18}
\bibinfo{author}{M.~Palu{\v{s}}}, \bibinfo{title}{Linked by Dynamics:
  Wavelet-Based Mutual Information Rate as a Connectivity Measure and
  Scale-Specific Networks}, \bibinfo{publisher}{Springer International
  Publishing}, \bibinfo{address}{Cham}, pp. \bibinfo{pages}{427--463}.
\bibitem[{Cha(2007)}]{cha07}
\bibinfo{author}{S.-H. Cha},
\newblock \bibinfo{title}{Comprehensive survey on distance/similarity measures
  between probability density functions},
\newblock \bibinfo{journal}{International Journal of Mathematical Models and
  Methods in Applied Sciences} \bibinfo{volume}{1} (\bibinfo{year}{2007})
  \bibinfo{pages}{300--307}.
\bibitem[{Deza and Deza(2009)}]{deza09}
\bibinfo{author}{M.~M. Deza}, \bibinfo{author}{E.~Deza},
\newblock \bibinfo{title}{Encyclopedia of distances},
\newblock in: \bibinfo{booktitle}{Encyclopedia of Distances},
  \bibinfo{publisher}{Springer}, \bibinfo{year}{2009}.
\bibitem[{Esling and Agon(2012)}]{esling12}
\bibinfo{author}{P.~Esling}, \bibinfo{author}{C.~Agon},
\newblock \bibinfo{title}{{Time-series data mining}},
\newblock \bibinfo{journal}{ACM Computing Surveys} \bibinfo{volume}{45}
  (\bibinfo{year}{2012}) \bibinfo{pages}{1--34}.
\bibitem[{Ferreira and Zhao(2016)}]{ferreira16}
\bibinfo{author}{L.~N. Ferreira}, \bibinfo{author}{L.~Zhao},
\newblock \bibinfo{title}{Time series clustering via community detection in
  networks},
\newblock \bibinfo{journal}{Information Sciences} \bibinfo{volume}{326}
  (\bibinfo{year}{2016}) \bibinfo{pages}{227 -- 242}.
\bibitem[{Barab{\'a}si and P{\'o}sfai(2016)}]{barabasi16}
\bibinfo{author}{A.~Barab{\'a}si}, \bibinfo{author}{M.~P{\'o}sfai},
  \bibinfo{title}{Network Science}, \bibinfo{publisher}{Cambridge University
  Press}, \bibinfo{year}{2016}.
\bibitem[{Boccaletti et~al.(2006)Boccaletti, Latora, Moreno, Chavez, and
  Hwang}]{boccaletti06}
\bibinfo{author}{S.~Boccaletti}, \bibinfo{author}{V.~Latora},
  \bibinfo{author}{Y.~Moreno}, \bibinfo{author}{M.~Chavez},
  \bibinfo{author}{D.-U. Hwang},
\newblock \bibinfo{title}{Complex networks: Structure and dynamics},
\newblock \bibinfo{journal}{Physics Reports} \bibinfo{volume}{424}
  (\bibinfo{year}{2006}) \bibinfo{pages}{175 -- 308}.
\bibitem[{da~F.~Costa et~al.(2007)da~F.~Costa, Rodrigues, Travieso, and
  Boas}]{costa07}
\bibinfo{author}{L.~da~F.~Costa}, \bibinfo{author}{F.~A. Rodrigues},
  \bibinfo{author}{G.~Travieso}, \bibinfo{author}{P.~R.~V. Boas},
\newblock \bibinfo{title}{Characterization of complex networks: A survey of
  measurements},
\newblock \bibinfo{journal}{Advances in Physics} \bibinfo{volume}{56}
  (\bibinfo{year}{2007}) \bibinfo{pages}{167--242}.
\bibitem[{Wiedermann et~al.(2017)Wiedermann, Donges, Kurths, and
  Donner}]{Wiedermann2017}
\bibinfo{author}{M.~Wiedermann}, \bibinfo{author}{J.~F. Donges},
  \bibinfo{author}{J.~Kurths}, \bibinfo{author}{R.~V. Donner},
\newblock \bibinfo{title}{Mapping and discrimination of networks in the
  complexity-entropy plane},
\newblock \bibinfo{journal}{Physical Review E} \bibinfo{volume}{96}
  (\bibinfo{year}{2017}) \bibinfo{pages}{042304}.
\bibitem[{Bialonski et~al.(2010)Bialonski, Horstmann, and
  Lehnertz}]{Bialonski2010}
\bibinfo{author}{S.~Bialonski}, \bibinfo{author}{M.-T. Horstmann},
  \bibinfo{author}{K.~Lehnertz},
\newblock \bibinfo{title}{From brain to earth and climate systems: Small-world
  interaction networks or not?},
\newblock \bibinfo{journal}{Chaos} \bibinfo{volume}{20} (\bibinfo{year}{2010})
  \bibinfo{pages}{013134}.
\bibitem[{Bialonski et~al.(2011)Bialonski, Wendler, and
  Lehnertz}]{Bialonski2011}
\bibinfo{author}{S.~Bialonski}, \bibinfo{author}{M.~Wendler},
  \bibinfo{author}{K.~Lehnertz},
\newblock \bibinfo{title}{Unraveling spurious properties of interaction
  networks with tailored random networks},
\newblock \bibinfo{journal}{PLOS ONE} \bibinfo{volume}{6}
  (\bibinfo{year}{2011}) \bibinfo{pages}{e22826}.
\bibitem[{Hlinka et~al.(2012)Hlinka, Hartman, and Paluš}]{Hlinka2012}
\bibinfo{author}{J.~Hlinka}, \bibinfo{author}{D.~Hartman},
  \bibinfo{author}{M.~Paluš},
\newblock \bibinfo{title}{Small-world topology of functional connectivity in
  randomly connected dynamical systems},
\newblock \bibinfo{journal}{Chaos} \bibinfo{volume}{22} (\bibinfo{year}{2012})
  \bibinfo{pages}{033107}.
\bibitem[{Wiedermann et~al.(2016)Wiedermann, Donges, Kurths, and
  Donner}]{Wiedermann2016}
\bibinfo{author}{M.~Wiedermann}, \bibinfo{author}{J.~F. Donges},
  \bibinfo{author}{J.~Kurths}, \bibinfo{author}{R.~V. Donner},
\newblock \bibinfo{title}{Spatial network surrogates for disentangling complex
  system structure from spatial embedding of nodes},
\newblock \bibinfo{journal}{Physical Review E} \bibinfo{volume}{93}
  (\bibinfo{year}{2016}) \bibinfo{pages}{042308}.
\bibitem[{Palu\v{s} et~al.(2011)Palu\v{s}, Hartman, Hlinka, and
  Vejmelka}]{palus11}
\bibinfo{author}{M.~Palu\v{s}}, \bibinfo{author}{D.~Hartman},
  \bibinfo{author}{J.~Hlinka}, \bibinfo{author}{M.~Vejmelka},
\newblock \bibinfo{title}{Discerning connectivity from dynamics in climate
  networks},
\newblock \bibinfo{journal}{Nonlinear Processes in Geophysics}
  \bibinfo{volume}{18} (\bibinfo{year}{2011}) \bibinfo{pages}{751--763}.
\bibitem[{Nigam and Baxter(2015)}]{nigam15}
\bibinfo{author}{S.~Nigam}, \bibinfo{author}{S.~Baxter},
\newblock \bibinfo{title}{General circulation of the atmosphere |
  teleconnections},
\newblock in: \bibinfo{editor}{G.~R. North}, \bibinfo{editor}{J.~Pyle},
  \bibinfo{editor}{F.~Zhang} (Eds.), \bibinfo{booktitle}{Encyclopedia of
  Atmospheric Sciences}, \bibinfo{publisher}{Academic Press},
  \bibinfo{address}{Oxford}, \bibinfo{edition}{2nd} edition,
  \bibinfo{year}{2015}, pp. \bibinfo{pages}{90 -- 109}.
\bibitem[{Zhou et~al.(2015)Zhou, Gozolchiani, Ashkenazy, and Havlin}]{dong15}
\bibinfo{author}{D.~Zhou}, \bibinfo{author}{A.~Gozolchiani},
  \bibinfo{author}{Y.~Ashkenazy}, \bibinfo{author}{S.~Havlin},
\newblock \bibinfo{title}{Teleconnection paths via climate network direct link
  detection},
\newblock \bibinfo{journal}{Physical Review Letters} \bibinfo{volume}{115}
  (\bibinfo{year}{2015}) \bibinfo{pages}{268501}.
\bibitem[{Alexander et~al.(2002)Alexander, Blad{\'e}, Newman, Lanzante, Lau,
  and Scott}]{alexander02}
\bibinfo{author}{M.~A. Alexander}, \bibinfo{author}{I.~Blad{\'e}},
  \bibinfo{author}{M.~Newman}, \bibinfo{author}{J.~R. Lanzante},
  \bibinfo{author}{N.-C. Lau}, \bibinfo{author}{J.~D. Scott},
\newblock \bibinfo{title}{{The Atmospheric Bridge: The Influence of ENSO
  Teleconnections on Air--Sea Interaction over the Global Oceans}},
\newblock \bibinfo{journal}{Journal of Climate} \bibinfo{volume}{15}
  (\bibinfo{year}{2002}) \bibinfo{pages}{2205--2231}.
\bibitem[{Boers et~al.(2019)Boers, Goswami, Rheinwalt, Bookhagen, Hoskins, and
  Kurths}]{boers19}
\bibinfo{author}{N.~Boers}, \bibinfo{author}{B.~Goswami},
  \bibinfo{author}{A.~Rheinwalt}, \bibinfo{author}{B.~Bookhagen},
  \bibinfo{author}{B.~Hoskins}, \bibinfo{author}{J.~Kurths},
\newblock \bibinfo{title}{Complex networks reveal global pattern of
  extreme-rainfall teleconnections},
\newblock \bibinfo{journal}{Nature}  (\bibinfo{year}{2019}).
\bibitem[{Yamasaki et~al.(2008)Yamasaki, Gozolchiani, and Havlin}]{yamasaki08}
\bibinfo{author}{K.~Yamasaki}, \bibinfo{author}{A.~Gozolchiani},
  \bibinfo{author}{S.~Havlin},
\newblock \bibinfo{title}{{Climate Networks around the Globe are Significantly
  Affected by El Ni\~no}},
\newblock \bibinfo{journal}{Physical Review Letters} \bibinfo{volume}{100}
  (\bibinfo{year}{2008}) \bibinfo{pages}{228501}.
\bibitem[{Kalnay et~al.(1996)Kalnay, Kanamitsu, Kistler, Collins, Deaven,
  Gandin, Iredell, Saha, White, Woollen, Zhu, Leetmaa, Reynolds, Chelliah,
  Ebisuzaki, Higgins, Janowiak, Mo, Ropelewski, Wang, Jenne, and
  Joseph}]{kalnay96}
\bibinfo{author}{E.~Kalnay}, \bibinfo{author}{M.~Kanamitsu},
  \bibinfo{author}{R.~Kistler}, \bibinfo{author}{W.~Collins},
  \bibinfo{author}{D.~Deaven}, \bibinfo{author}{L.~Gandin},
  \bibinfo{author}{M.~Iredell}, \bibinfo{author}{S.~Saha},
  \bibinfo{author}{G.~White}, \bibinfo{author}{J.~Woollen},
  \bibinfo{author}{Y.~Zhu}, \bibinfo{author}{A.~Leetmaa},
  \bibinfo{author}{R.~Reynolds}, \bibinfo{author}{M.~Chelliah},
  \bibinfo{author}{W.~Ebisuzaki}, \bibinfo{author}{W.~Higgins},
  \bibinfo{author}{J.~Janowiak}, \bibinfo{author}{K.~C. Mo},
  \bibinfo{author}{C.~Ropelewski}, \bibinfo{author}{J.~Wang},
  \bibinfo{author}{R.~Jenne}, \bibinfo{author}{D.~Joseph},
\newblock \bibinfo{title}{{The NCEP/NCAR 40-Year Reanalysis Project}},
\newblock \bibinfo{journal}{Bulletin of the American Meteorological Society}
  \bibinfo{volume}{77} (\bibinfo{year}{1996}) \bibinfo{pages}{437--471}.
\bibitem[{Berezin et~al.(2012)Berezin, Gozolchiani, Guez, and
  Havlin}]{berezin12}
\bibinfo{author}{Y.~Berezin}, \bibinfo{author}{A.~Gozolchiani},
  \bibinfo{author}{O.~Guez}, \bibinfo{author}{S.~Havlin},
\newblock \bibinfo{title}{Stability of climate networks with time},
\newblock \bibinfo{journal}{Scientific Reports} \bibinfo{volume}{2}
  (\bibinfo{year}{2012}) \bibinfo{pages}{666}.
\bibitem[{Deza et~al.(2015)Deza, Barreiro, and Masoller}]{deza15}
\bibinfo{author}{J.~I. Deza}, \bibinfo{author}{M.~Barreiro},
  \bibinfo{author}{C.~Masoller},
\newblock \bibinfo{title}{Assessing the direction of climate interactions by
  means of complex networks and information theoretic tools},
\newblock \bibinfo{journal}{Chaos} \bibinfo{volume}{25} (\bibinfo{year}{2015})
  \bibinfo{pages}{033105}.
\bibitem[{Bracco et~al.(2018)Bracco, Falasca, Nenes, Fountalis, and
  Dovrolis}]{bracco18}
\bibinfo{author}{A.~Bracco}, \bibinfo{author}{F.~Falasca},
  \bibinfo{author}{A.~Nenes}, \bibinfo{author}{I.~Fountalis},
  \bibinfo{author}{C.~Dovrolis},
\newblock \bibinfo{title}{Advancing climate science with knowledge-discovery
  through data mining},
\newblock \bibinfo{journal}{npj Climate and Atmospheric Science}
  \bibinfo{volume}{1} (\bibinfo{year}{2018}) \bibinfo{pages}{20174}.
\bibitem[{Pelan et~al.(2011)Pelan, Steinhaeuser, Chawla, de~Alwis~Pitts, and
  Ganguly}]{pelan11}
\bibinfo{author}{A.~Pelan}, \bibinfo{author}{K.~Steinhaeuser},
  \bibinfo{author}{N.~V. Chawla}, \bibinfo{author}{D.~A. de~Alwis~Pitts},
  \bibinfo{author}{A.~R. Ganguly},
\newblock \bibinfo{title}{Empirical comparison of correlation measures and
  pruning levels in complex networks representing the global climate system},
\newblock in: \bibinfo{booktitle}{2011 IEEE Symposium on Computational
  Intelligence and Data Mining (CIDM)}, pp. \bibinfo{pages}{239--245}.
\bibitem[{Wolf et~al.(2020)Wolf, Bauer, Boers, and Donner}]{Wolf2020}
\bibinfo{author}{F.~Wolf}, \bibinfo{author}{J.~Bauer},
  \bibinfo{author}{N.~Boers}, \bibinfo{author}{R.~V. Donner},
\newblock \bibinfo{title}{{Event synchrony measures for functional climate
  network analysis: A case study on South American rainfall dynamics}},
\newblock \bibinfo{journal}{Chaos} \bibinfo{volume}{30} (\bibinfo{year}{2020})
  \bibinfo{pages}{033102}.
\bibitem[{Yamasaki et~al.(2009)Yamasaki, Gozolchiani, and
  Havlin}]{Yamasaki2009}
\bibinfo{author}{K.~Yamasaki}, \bibinfo{author}{A.~Gozolchiani},
  \bibinfo{author}{S.~Havlin},
\newblock \bibinfo{title}{{Climate Networks Based on Phase Synchronization
  Analysis Track El-Niño}},
\newblock \bibinfo{journal}{Progress of Theoretical Physics Supplement}
  \bibinfo{volume}{179} (\bibinfo{year}{2009}) \bibinfo{pages}{178--188}.
\bibitem[{Berndt and Clifford(1994)}]{berndt94}
\bibinfo{author}{D.~J. Berndt}, \bibinfo{author}{J.~Clifford},
\newblock \bibinfo{title}{Using dynamic time warping to find patterns in time
  series.},
\newblock in: \bibinfo{booktitle}{KDD Workshop}, \bibinfo{publisher}{AAAI
  Press}, \bibinfo{year}{1994}, pp. \bibinfo{pages}{359--370}.
\bibitem[{Batista et~al.(2011)Batista, Wang, and Keogh}]{batista11}
\bibinfo{author}{G.~E. Batista}, \bibinfo{author}{X.~Wang},
  \bibinfo{author}{E.~J. Keogh},
\newblock \bibinfo{title}{A complexity-invariant distance measure for time
  series},
\newblock in: \bibinfo{booktitle}{Proceedings of the 2011 SIAM International
  Conference on Data Mining}, \bibinfo{organization}{SIAM}, pp.
  \bibinfo{pages}{699--710}.
\bibitem[{Frentzos et~al.(2007)Frentzos, Gratsias, and
  Theodoridis}]{frentzos07}
\bibinfo{author}{E.~Frentzos}, \bibinfo{author}{K.~Gratsias},
  \bibinfo{author}{Y.~Theodoridis},
\newblock \bibinfo{title}{Index-based most similar trajectory search},
\newblock in: \bibinfo{booktitle}{2007 IEEE 23rd International Conference on
  Data Engineering}, pp. \bibinfo{pages}{816--825}.
\bibitem[{M{\"o}ller-Levet et~al.(2003)M{\"o}ller-Levet, Klawonn, Cho, and
  Wolkenhauer}]{moller03}
\bibinfo{author}{C.~S. M{\"o}ller-Levet}, \bibinfo{author}{F.~Klawonn},
  \bibinfo{author}{K.-H. Cho}, \bibinfo{author}{O.~Wolkenhauer},
  \bibinfo{title}{Fuzzy Clustering of Short Time-Series and Unevenly
  Distributed Sampling Points}, \bibinfo{publisher}{Springer Berlin
  Heidelberg}, \bibinfo{address}{Berlin, Heidelberg}, pp.
  \bibinfo{pages}{330--340}.
\bibitem[{Chen et~al.(2005)Chen, \"{O}zsu, and Oria}]{chen05}
\bibinfo{author}{L.~Chen}, \bibinfo{author}{M.~T. \"{O}zsu},
  \bibinfo{author}{V.~Oria},
\newblock \bibinfo{title}{Robust and fast similarity search for moving object
  trajectories},
\newblock in: \bibinfo{booktitle}{Proceedings of the 2005 ACM SIGMOD
  International Conference on Management of Data}, SIGMOD '05,
  \bibinfo{publisher}{ACM}, \bibinfo{address}{New York, NY, USA},
  \bibinfo{year}{2005}, pp. \bibinfo{pages}{491--502}.
\bibitem[{Vlachos et~al.(2002)Vlachos, Kollios, and Gunopulos}]{vlachos02}
\bibinfo{author}{M.~Vlachos}, \bibinfo{author}{G.~Kollios},
  \bibinfo{author}{D.~Gunopulos},
\newblock \bibinfo{title}{Discovering similar multidimensional trajectories},
\newblock in: \bibinfo{booktitle}{Data Engineering, 2002. Proceedings. 18th
  International Conference on}, pp. \bibinfo{pages}{673--684}.
\bibitem[{Chouakria and Nagabhushan(2007)}]{chouakria07}
\bibinfo{author}{A.~D. Chouakria}, \bibinfo{author}{P.~N. Nagabhushan},
\newblock \bibinfo{title}{Adaptive dissimilarity index for measuring time
  series proximity},
\newblock \bibinfo{journal}{Advances in Data Analysis and Classification}
  \bibinfo{volume}{1} (\bibinfo{year}{2007}) \bibinfo{pages}{5--21}.
\bibitem[{Meil{\u{a}}(2003)}]{meila03}
\bibinfo{author}{M.~Meil{\u{a}}}, \bibinfo{title}{Comparing Clusterings by the
  Variation of Information}, \bibinfo{publisher}{Springer Berlin Heidelberg},
  \bibinfo{address}{Berlin, Heidelberg}, pp. \bibinfo{pages}{173--187}.
\bibitem[{Reshef et~al.(2011)Reshef, Reshef, Finucane, Grossman, McVean,
  Turnbaugh, Lander, Mitzenmacher, and Sabeti}]{reshef11}
\bibinfo{author}{D.~N. Reshef}, \bibinfo{author}{Y.~A. Reshef},
  \bibinfo{author}{H.~K. Finucane}, \bibinfo{author}{S.~R. Grossman},
  \bibinfo{author}{G.~McVean}, \bibinfo{author}{P.~J. Turnbaugh},
  \bibinfo{author}{E.~S. Lander}, \bibinfo{author}{M.~Mitzenmacher},
  \bibinfo{author}{P.~C. Sabeti},
\newblock \bibinfo{title}{Detecting novel associations in large data sets},
\newblock \bibinfo{journal}{Science} \bibinfo{volume}{334}
  (\bibinfo{year}{2011}) \bibinfo{pages}{1518--1524}.
\bibitem[{Agrawal et~al.(1993)Agrawal, Faloutsos, and Swami}]{agrawal93}
\bibinfo{author}{R.~Agrawal}, \bibinfo{author}{C.~Faloutsos},
  \bibinfo{author}{A.~Swami}, \bibinfo{title}{Efficient similarity search in
  sequence databases}, \bibinfo{publisher}{Springer Berlin Heidelberg},
  \bibinfo{address}{Berlin, Heidelberg}, pp. \bibinfo{pages}{69--84}.
\bibitem[{Galeano and Pe{\~n}a(2000)}]{galeano00}
\bibinfo{author}{P.~Galeano}, \bibinfo{author}{D.~P. Pe{\~n}a},
\newblock \bibinfo{title}{Multivariate analysis in vector time series},
\newblock \bibinfo{journal}{Resenhas do Instituto de Matem{\'a}tica e
  Estat\'{i}stica da Universidade de S{\~a}o Paulo} \bibinfo{volume}{4}
  (\bibinfo{year}{2000}) \bibinfo{pages}{383--403}.
\bibitem[{de~Lucas(2003)}]{casado03}
\bibinfo{author}{D.~C. de~Lucas}, \bibinfo{title}{Classification Techniques for
  Time Series and Functional Data}, Ph.D. thesis, Universidad Carlos III de
  Madrid, \bibinfo{year}{2003}.
\bibitem[{Caiado et~al.(2006)Caiado, Crato, and Pe{\~n}a}]{caiado06}
\bibinfo{author}{J.~Caiado}, \bibinfo{author}{N.~Crato},
  \bibinfo{author}{D.~Pe{\~n}a},
\newblock \bibinfo{title}{A periodogram-based metric for time series
  classification},
\newblock \bibinfo{journal}{Computational Statistics \& Data Analysis}
  \bibinfo{volume}{50} (\bibinfo{year}{2006}) \bibinfo{pages}{2668 -- 2684}.
\bibitem[{Cilibrasi and Vitanyi(2005)}]{cilibrasi05}
\bibinfo{author}{R.~Cilibrasi}, \bibinfo{author}{P.~M.~B. Vitanyi},
\newblock \bibinfo{title}{Clustering by compression},
\newblock \bibinfo{journal}{IEEE Transactions on Information Theory}
  \bibinfo{volume}{51} (\bibinfo{year}{2005}) \bibinfo{pages}{1523--1545}.
\bibitem[{Keogh et~al.(2007)Keogh, Lonardi, Ratanamahatana, Wei, Lee, and
  Handley}]{keogh07}
\bibinfo{author}{E.~Keogh}, \bibinfo{author}{S.~Lonardi},
  \bibinfo{author}{C.~A. Ratanamahatana}, \bibinfo{author}{L.~Wei},
  \bibinfo{author}{S.-H. Lee}, \bibinfo{author}{J.~Handley},
\newblock \bibinfo{title}{Compression-based data mining of sequential data},
\newblock \bibinfo{journal}{Data Mining and Knowledge Discovery}
  \bibinfo{volume}{14} (\bibinfo{year}{2007}) \bibinfo{pages}{99--129}.
\bibitem[{Kendall and Stuart(1983)}]{kendall83}
\bibinfo{author}{M.~Kendall}, \bibinfo{author}{A.~Stuart}, \bibinfo{title}{The
  advanced theory of statistics, vol. 3}, \bibinfo{year}{1983}.
\bibitem[{Bhattacharyya(1946)}]{bhattacharyya46}
\bibinfo{author}{A.~Bhattacharyya},
\newblock \bibinfo{title}{On a measure of divergence between two multinomial
  populations},
\newblock \bibinfo{journal}{Sankhy{\=a}: The Indian Journal of Statistics
  (1933-1960)} \bibinfo{volume}{7} (\bibinfo{year}{1946})
  \bibinfo{pages}{401--406}.
\bibitem[{Dice(1945)}]{dice45}
\bibinfo{author}{L.~R. Dice},
\newblock \bibinfo{title}{Measures of the amount of ecologic association
  between species},
\newblock \bibinfo{journal}{Ecology} \bibinfo{volume}{26}
  (\bibinfo{year}{1945}) \bibinfo{pages}{297--302}.
\bibitem[{Gower(1971)}]{gower71}
\bibinfo{author}{J.~C. Gower},
\newblock \bibinfo{title}{A general coefficient of similarity and some of its
  properties},
\newblock \bibinfo{journal}{Biometrics} \bibinfo{volume}{27}
  (\bibinfo{year}{1971}) \bibinfo{pages}{857--871}.
\bibitem[{S{\o}rensen(1948)}]{sorensen48}
\bibinfo{author}{T.~S{\o}rensen}, \bibinfo{title}{A Method of Establishing
  Groups of Equal Amplitude in Plant Sociology Based on Similarity of Species
  Content and Its Application to Analyses of the Vegetation on Danish Commons},
  Biologiske Skrifter // Det Kongelige Danske Videnskabernes Selskab,
  \bibinfo{publisher}{I kommission hos E. Munksgaard}, \bibinfo{year}{1948}.
\bibitem[{Tanimoto(1958)}]{tanimoto58}
\bibinfo{author}{T.~Tanimoto}, \bibinfo{title}{An Elementary Mathematical
  Theory of Classification and Prediction}, \bibinfo{publisher}{International
  Business Machines Corporation}, \bibinfo{year}{1958}.
\bibitem[{Levenshtein(1966)}]{levenshtein66}
\bibinfo{author}{V.~I. Levenshtein},
\newblock \bibinfo{title}{{Binary Codes Capable of Correcting Deletions,
  Insertions and Reversals}},
\newblock \bibinfo{journal}{Soviet Physics Doklady} \bibinfo{volume}{10}
  (\bibinfo{year}{1966}) \bibinfo{pages}{707--710}.
\bibitem[{Humphries and Gurney(2008)}]{humphries08}
\bibinfo{author}{M.~D. Humphries}, \bibinfo{author}{K.~Gurney},
\newblock \bibinfo{title}{Network `small-world-ness': A quantitative method for
  determining canonical network equivalence},
\newblock \bibinfo{journal}{PLOS ONE} \bibinfo{volume}{3}
  (\bibinfo{year}{2008}) \bibinfo{pages}{1--10}.
\bibitem[{Glantz et~al.(1991)Glantz, Katz, Nicholls et~al.}]{glantz91}
\bibinfo{author}{M.~H. Glantz}, \bibinfo{author}{R.~W. Katz},
  \bibinfo{author}{N.~Nicholls}, et~al., \bibinfo{title}{Teleconnections
  linking worldwide climate anomalies}, volume \bibinfo{volume}{535},
  \bibinfo{publisher}{Cambridge University Press Cambridge},
  \bibinfo{year}{1991}.
\bibitem[{Hlinka et~al.(2017)Hlinka, Hartman, Jajcay, Tome{\v c}ek, Tint{\v
  e}ra, and Palu{\v s}}]{hlinka17}
\bibinfo{author}{J.~Hlinka}, \bibinfo{author}{D.~Hartman},
  \bibinfo{author}{N.~Jajcay}, \bibinfo{author}{D.~Tome{\v c}ek},
  \bibinfo{author}{J.~Tint{\v e}ra}, \bibinfo{author}{M.~Palu{\v s}},
\newblock \bibinfo{title}{Small-world bias of correlation networks: From brain
  to climate},
\newblock \bibinfo{journal}{Chaos: An Interdisciplinary Journal of Nonlinear
  Science} \bibinfo{volume}{27} (\bibinfo{year}{2017}) \bibinfo{pages}{035812}.
\bibitem[{Amaral et~al.(2000)Amaral, Scala, Barth{\'e}l{\'e}my, and
  Stanley}]{amaral20}
\bibinfo{author}{L.~A.~N. Amaral}, \bibinfo{author}{A.~Scala},
  \bibinfo{author}{M.~Barth{\'e}l{\'e}my}, \bibinfo{author}{H.~E. Stanley},
\newblock \bibinfo{title}{Classes of small-world networks},
\newblock \bibinfo{journal}{Proceedings of the National Academy of Sciences}
  \bibinfo{volume}{97} (\bibinfo{year}{2000}) \bibinfo{pages}{11149--11152}.
\bibitem[{Radebach et~al.(2013)Radebach, Donner, Runge, Donges, and
  Kurths}]{Radebach2013}
\bibinfo{author}{A.~Radebach}, \bibinfo{author}{R.~V. Donner},
  \bibinfo{author}{J.~Runge}, \bibinfo{author}{J.~F. Donges},
  \bibinfo{author}{J.~Kurths},
\newblock \bibinfo{title}{{Disentangling different types of El Ni\~no episodes
  by evolving climate network analysis}},
\newblock \bibinfo{journal}{Physical Review E} \bibinfo{volume}{88}
  (\bibinfo{year}{2013}) \bibinfo{pages}{052807}.
\bibitem[{Liu and Alexander(2007)}]{liu2007atmospheric}
\bibinfo{author}{Z.~Liu}, \bibinfo{author}{M.~Alexander},
\newblock \bibinfo{title}{Atmospheric bridge, oceanic tunnel, and global
  climatic teleconnections},
\newblock \bibinfo{journal}{Reviews of Geophysics} \bibinfo{volume}{45}
  (\bibinfo{year}{2007}).
\bibitem[{Picaut et~al.(1997)Picaut, Masia, and Du~Penhoat}]{picaut97}
\bibinfo{author}{J.~Picaut}, \bibinfo{author}{F.~Masia},
  \bibinfo{author}{Y.~Du~Penhoat},
\newblock \bibinfo{title}{{An advective-reflective conceptual model for the
  oscillatory nature of the ENSO}},
\newblock \bibinfo{journal}{Science} \bibinfo{volume}{277}
  (\bibinfo{year}{1997}) \bibinfo{pages}{663--666}.
\bibitem[{Gong and Wang(1999)}]{gong1999definition}
\bibinfo{author}{D.~Gong}, \bibinfo{author}{S.~Wang},
\newblock \bibinfo{title}{{Definition of Antarctic oscillation index}},
\newblock \bibinfo{journal}{Geophysical Research Letters} \bibinfo{volume}{26}
  (\bibinfo{year}{1999}) \bibinfo{pages}{459--462}.
\bibitem[{Mo and Higgins(1998)}]{mo1998pacific}
\bibinfo{author}{K.~C. Mo}, \bibinfo{author}{R.~W. Higgins},
\newblock \bibinfo{title}{{The Pacific--South American modes and tropical
  convection during the Southern Hemisphere winter}},
\newblock \bibinfo{journal}{Monthly Weather Review} \bibinfo{volume}{126}
  (\bibinfo{year}{1998}) \bibinfo{pages}{1581--1596}.
\bibitem[{Mo and Paegle(2001)}]{mo01}
\bibinfo{author}{K.~C. Mo}, \bibinfo{author}{J.~N. Paegle},
\newblock \bibinfo{title}{The pacific--south american modes and their
  downstream effects},
\newblock \bibinfo{journal}{International Journal of Climatology}
  \bibinfo{volume}{21} (\bibinfo{year}{2001}) \bibinfo{pages}{1211--1229}.
\bibitem[{Renwick and Revell(1999)}]{renwick99}
\bibinfo{author}{J.~A. Renwick}, \bibinfo{author}{M.~J. Revell},
\newblock \bibinfo{title}{Blocking over the south pacific and rossby wave
  propagation},
\newblock \bibinfo{journal}{Monthly Weather Review} \bibinfo{volume}{127}
  (\bibinfo{year}{1999}) \bibinfo{pages}{2233--2247}.
\bibitem[{Thompson and Wallace(2000)}]{thompson2000annular}
\bibinfo{author}{D.~W. Thompson}, \bibinfo{author}{J.~M. Wallace},
\newblock \bibinfo{title}{Annular modes in the extratropical circulation. part
  i: Month-to-month variability},
\newblock \bibinfo{journal}{Journal of Climate} \bibinfo{volume}{13}
  (\bibinfo{year}{2000}) \bibinfo{pages}{1000--1016}.
\bibitem[{Kidson(1988)}]{kidson1988indices}
\bibinfo{author}{J.~W. Kidson},
\newblock \bibinfo{title}{{Indices of the Southern Hemisphere zonal wind}},
\newblock \bibinfo{journal}{Journal of Climate} \bibinfo{volume}{1}
  (\bibinfo{year}{1988}) \bibinfo{pages}{183--194}.
\bibitem[{Vasconcellos and Cavalcanti(2010)}]{vasconcellos2010extreme}
\bibinfo{author}{F.~C. Vasconcellos}, \bibinfo{author}{I.~F. Cavalcanti},
\newblock \bibinfo{title}{{Extreme precipitation over Southeastern Brazil in
  the austral summer and relations with the Southern Hemisphere annular mode}},
\newblock \bibinfo{journal}{Atmospheric Science Letters} \bibinfo{volume}{11}
  (\bibinfo{year}{2010}) \bibinfo{pages}{21--26}.
\bibitem[{Aceituno(1988)}]{aceituno88}
\bibinfo{author}{P.~Aceituno},
\newblock \bibinfo{title}{{On the functioning of the Southern Oscillation in
  the South American sector. Part I: Surface climate}},
\newblock \bibinfo{journal}{Monthly Weather Review} \bibinfo{volume}{116}
  (\bibinfo{year}{1988}) \bibinfo{pages}{505--524}.
\bibitem[{Chan et~al.(2008)Chan, Behera, and Yamagata}]{chan2008indian}
\bibinfo{author}{S.~C. Chan}, \bibinfo{author}{S.~K. Behera},
  \bibinfo{author}{T.~Yamagata},
\newblock \bibinfo{title}{{Indian Ocean dipole influence on South American
  rainfall}},
\newblock \bibinfo{journal}{Geophysical Research Letters} \bibinfo{volume}{35}
  (\bibinfo{year}{2008}).
\bibitem[{Saji and Yamagata(2003)}]{saji2003possible}
\bibinfo{author}{N.~Saji}, \bibinfo{author}{T.~Yamagata},
\newblock \bibinfo{title}{Possible impacts of indian ocean dipole mode events
  on global climate},
\newblock \bibinfo{journal}{Climate Research} \bibinfo{volume}{25}
  (\bibinfo{year}{2003}) \bibinfo{pages}{151--169}.
\bibitem[{Yamagata et~al.(2004)Yamagata, Behera, Luo, Masson, Jury, and
  Rao}]{yamagata2004coupled}
\bibinfo{author}{T.~Yamagata}, \bibinfo{author}{S.~K. Behera},
  \bibinfo{author}{J.-J. Luo}, \bibinfo{author}{S.~Masson},
  \bibinfo{author}{M.~R. Jury}, \bibinfo{author}{S.~A. Rao},
\newblock \bibinfo{title}{{Coupled ocean-atmosphere variability in the tropical
  Indian Ocean}},
\newblock \bibinfo{journal}{Earth’s Climate: The Ocean--Atmosphere
  Interaction, Geophysical Monographs} \bibinfo{volume}{147}
  (\bibinfo{year}{2004}) \bibinfo{pages}{189--212}.
\bibitem[{Behera et~al.(2005)Behera, Luo, Masson, Delecluse, Gualdi, Navarra,
  and Yamagata}]{behera2005paramount}
\bibinfo{author}{S.~K. Behera}, \bibinfo{author}{J.-J. Luo},
  \bibinfo{author}{S.~Masson}, \bibinfo{author}{P.~Delecluse},
  \bibinfo{author}{S.~Gualdi}, \bibinfo{author}{A.~Navarra},
  \bibinfo{author}{T.~Yamagata},
\newblock \bibinfo{title}{{Paramount impact of the Indian Ocean dipole on the
  East African short rains: A CGCM study}},
\newblock \bibinfo{journal}{Journal of Climate} \bibinfo{volume}{18}
  (\bibinfo{year}{2005}) \bibinfo{pages}{4514--4530}.
\bibitem[{Saji et~al.(1999)Saji, Goswami, Vinayachandran, and
  Yamagata}]{saji1999dipole}
\bibinfo{author}{N.~Saji}, \bibinfo{author}{B.~Goswami},
  \bibinfo{author}{P.~Vinayachandran}, \bibinfo{author}{T.~Yamagata},
\newblock \bibinfo{title}{{A dipole mode in the tropical Indian Ocean}},
\newblock \bibinfo{journal}{Nature} \bibinfo{volume}{401}
  (\bibinfo{year}{1999}) \bibinfo{pages}{360}.
\bibitem[{{R Core Team}(2017)}]{rstats}
\bibinfo{author}{{R Core Team}}, \bibinfo{title}{R: A Language and Environment
  for Statistical Computing}, \bibinfo{organization}{R Foundation for
  Statistical Computing}, \bibinfo{address}{Vienna, Austria},
  \bibinfo{year}{2017}.
\bibitem[{Csardi and Nepusz(2006)}]{igraph06}
\bibinfo{author}{G.~Csardi}, \bibinfo{author}{T.~Nepusz},
\newblock \bibinfo{title}{The igraph software package for complex network
  research},
\newblock \bibinfo{journal}{InterJournal} \bibinfo{volume}{Complex Systems}
  (\bibinfo{year}{2006}) \bibinfo{pages}{1695}.
\bibitem[{Montero and Vilar(2014)}]{tsclust14}
\bibinfo{author}{P.~Montero}, \bibinfo{author}{J.~Vilar},
\newblock \bibinfo{title}{{TSclust: An R Package for Time Series Clustering}},
\newblock \bibinfo{journal}{Journal of Statistical Software}
  \bibinfo{volume}{62} (\bibinfo{year}{2014}) \bibinfo{pages}{1--43}.
\bibitem[{Mori et~al.(2016)Mori, Mendiburu, and Lozano}]{mori16}
\bibinfo{author}{U.~Mori}, \bibinfo{author}{A.~Mendiburu},
  \bibinfo{author}{J.~A. Lozano},
\newblock \bibinfo{title}{{Distance Measures for Time Series in R: The {TSdist}
  Package}},
\newblock \bibinfo{journal}{R Journal} \bibinfo{volume}{8}
  (\bibinfo{year}{2016}) \bibinfo{pages}{451--459}.

\end{thebibliography}

\newpage
\appendix
\singlespacing

\section{Definitions of time series distance functions} 
\label{appendix:ts_ds}


\begin{table}[!ht]
  \centering
  \footnotesize
  \caption{Time series distance functions used in this work} 
  \label{tab:measures_definitions_1}
  \begin{tabular}[t]{llc}
    \toprule
    Distance & Equation  & Ref. \\ 
    \midrule        
    
    \rowcolor{gray!7}
    Avg($L_1$, $L_\infty$) (avgL1LInf) & $d_{ALL}(P,Q) = (\sum_{i}^{t}|P_i-Q_i| + \max{|P_i-Q_i|})/2$ & \cite{cha07} \\ [12px]

    Bhattacharyya & $d_{BH}(P,Q) = - \ln \sum_{i}^{t} \sqrt{P_iQ_i}$ & \cite{bhattacharyya46} \\ [8px]

    \rowcolor{gray!7}
    Dice & $d_{DC}(P,Q) = \displaystyle\frac{\sum_{i}^{t}(P_i-Q_i)^2}{\sum_{i}^{t}P_i^2 + \sum_{i}^{t}Q_i^2} $ & \cite{dice45} \\ [3ex]


    Gower & $d_{GW}(P,Q) = \displaystyle\frac{1}{t}\sum_{i}^{t}|P_i-Q_i| $ & \cite{gower71} \\[3ex]

    \rowcolor{gray!7}
    Jaccard & $d_{JC}(P,Q) = \displaystyle\frac{\sum_{i}^{t}(P_i-Q_i)^2}{\sum_{i}^{t}P_i^2 + \sum_{i}^{t}Q_i^2 - \sum_{i}^{t}P_i Q_i} $ & \cite{deza09} \\ [15px]

    Kulczynski & $d_{KS}(P,Q) = \displaystyle\frac{\sum_{i}^{t}|P_i-Q_i|}{\sum_{i}^{t}\min{(P_i,Q_i)}} $ & \cite{deza09} \\ [3ex]

    \rowcolor{gray!7}
    Lorentzian & $d_{LO}(P,Q) = \displaystyle\sum_{i}^{t} \ln(1+|P_i-Q_i|) $ & \cite{deza09} \\ [3ex]

    \pbox{30cm}{Mutual Information\textsuperscript{a} (MI)} 
    & $d_{MI}(X,Y) = H(X) + H(Y) - 2 MI(X,Y) $ & \cite{meila03} \\ [5px]
    & $ H(X) = - \sum_{i}^{t} P_i \log P_i $ & \\ [5px]
    & $MI(X,Y) = \displaystyle \sum_{i}^{t} \sum_{j}^{t} p(X_i,Y_j) \log \frac{p(X_i,Y_j)}{P_i Q_j}$ & 
    \\ [12px]
    
    \rowcolor{gray!7}
    \pbox{22cm}{ Maximal Information\\ Coefficient\textsuperscript{b} (MIC)}
    
      & $d_{MIC}(X,Y) = 1 - MIC(X,Y)$ & \cite{reshef11} \\ [5px]
      \rowcolor{gray!7}
      & $MIC(X,Y) = \displaystyle \max_{x,y < t^{\alpha}}(M_{xy})$ & \\ [12px]
      \rowcolor{gray!7}
      & $M_{xy} = \displaystyle \frac{\max(MI(G_{xy}))}{\log(\min(x,y))}$ & 
      \\ [12px]

    S\o rensen & $d_{SO}(P,Q) = \displaystyle\frac{\sum_{i}^{t}|P_i-Q_i|}{\sum_{i}^{t}(P_i+Q_i)} $ & \cite{sorensen48} \\ [3ex]


    \rowcolor{gray!7}
    \pbox{20cm}{Squared Euclidean\\(sqdEuclidean)} & $d_{SED}(P,Q) = \displaystyle\sum_{i}^{t}(P_i-Q_i)^2 $ & \cite{deza09} \\ [18px]

    Tanimoto & $d_{TM}(P,Q) = \displaystyle\frac{\sum_{i}^{t}(\max{(P_i,Q_i)}-\min{(P_i,Q_i)})}{ \sum_{i}^{t}\max{(P_i,Q_i)} } $ & \cite{tanimoto58} \\ [3ex]

    \rowcolor{gray!7}
    Wave Hedges & $d_{WH}(P,Q) = \displaystyle\sum_{i}^{t}\frac{|P_i-Q_i|}{\max{(P_i,Q_i)}} $ & \cite{cha07} \\ [3ex]

    \bottomrule
  \end{tabular}
  \begin{flushleft} 
        \item Given two equally sampled time series $X$ and $Y$, both with the same length $T$, we represent the respective probability distributions as $p(X) = P$ and $p(Y) = Q$;
        \item[\textsuperscript{a}]$p(X_i,Y_j)$ is the joint probability distribution function of $X_i$ and $Y_j$.
        \item[\textsuperscript{b}]$MI(G_{xy})$ is the MI of the probability distributions induced on the boxes of an $x-y$ grid $G_{xy}$. $\alpha = 0.6$ \cite{reshef11}.
  \end{flushleft}
\end{table}

\begin{table}[!ht]
  \centering
  \footnotesize
  \caption{Time series distance functions used in this work} 
  \label{tab:measures_definitions_2}
  \begin{tabular}[t]{llc}
    \toprule
    Distance & Equation  & Ref.            \\
    \midrule        
    
    \rowcolor{gray!7}
    \pbox{20cm}{Euclidean} & 
      $d_{\text{ED}}(X,Y) = \sqrt{\sum_{t=1}^{T}(X_t-Y_t)^2} $ & 
      \cite{deza09} \\ [8px]

    Manhattan & 
      $d_{\text{MH}}(X,Y) = \sum_{t=1}^{T}|X_t-Y_t| $ & 
      \cite{deza09} \\ [8px]


    \rowcolor{gray!7}
    \pbox{22cm}{Complexity\\Invariant\\(CID)} &
    $d_{\text{CID}}(X,Y) = cf(X,Y) \cdot d_{\text{ED}}(X,Y)$ & \cite{batista11}\\ [10px]
    \rowcolor{gray!7}
     & $cf(X,Y) = \frac{\max(ce(X),ce(Y))}{\min(ce(X),ce(Y))}$ & \\ [10px]
    \rowcolor{gray!7}
     & $ce(X) = \sqrt{\sum_{t=1}^{T-1}(X_t - X_{t+1})^2}$ & \\ [10px]
  
    \pbox{22cm}{Dynamic\\time\\warping\textsuperscript{a}\\(DTW)} & $d_{\text{DTW}}(X,Y) = dtw(i=T,j=T)$ & \cite{berndt94} \\ [8px]
    & $ 
    dtw(i,j) = 
        \begin{cases}
          \infty & \text{, if } i=0 \text{ xor } j=0\\
          0      & \text{, if } i=j=0\\
          |X_i-Y_j| + \min
        \begin{cases}
          dtw(i-1,j)  \\
          dtw(i,j-1)  \\
          dtw(i-1,j-1)  \\
        \end{cases}  & \text{, otherwise}\\
        \end{cases} 
    $ & \\ [10px]

    
    \rowcolor{gray!7}
    Correlation\textsuperscript{b} &  $d_{\text{COR}}(X,Y) = 1 - |\rho_{XY}|$ & \\ [8px]

      \rowcolor{gray!7}
      & $\rho_{XY}  = \displaystyle \frac{\sum_{t=1}^{T}(X_t-\bar{X})(Y_t-\bar{Y})}{\sqrt{\sum_{t=1}^{T}(X_t-\bar{X})^2 } \cdot \sqrt{\sum_{t=1}^{T}(Y_t-\bar{Y})^2}}$ & 
      \cite{deza09} \\ [22px]

     Cross-correlation\textsuperscript{b}\\(crossCor) &  
      $d_{\text{CCOR}}(X,Y) = \displaystyle 1 - \bigg(\max_{\tau \in [-\tau_{max},\tau_{max}]}|\rho_{XY_{\tau}}|\bigg) $ & 
      \cite{deza09} \\ [15px]
     & $\rho_{XY_{\tau}}  = \displaystyle \frac{\sum_{t=1}^{T}(X_t-\bar{X})(Y_{t-\tau}-\bar{Y})}{\sqrt{\sum_{t=1}^{T}(X_t-\bar{X})^2 } \cdot \sqrt{\sum_{t=1}^{T}(Y_{t-\tau}-\bar{Y})^2}}$  & \\ [25px]
    
    \rowcolor{gray!7}
    \pbox{22cm}{Autocorrelation\\Coefficients\textsuperscript{b}\\(ACF) \\ Partial\\Autocorrelation\\Coefficients\\(PACF)} & 
      $d_{\text{ACF}}(X,Y) = \displaystyle \sqrt{\sum_{\tau = -\tau_{max}}^{\tau_{max}}(\rho_{XX_{\tau}} - \rho_{YY_{\tau}})^2} $ & 
      \cite{galeano00} \\ [12px]

    \bottomrule
  \end{tabular}
  \begin{flushleft} 
        \item $X$ and $Y$ are two time series with the same length $T$.
        \item[\textsuperscript{a}] Definition using dynamic programming. The total distance follows the recurrence relation for $dtw(i,j)$, that is cumulative distance for each point.
        \item[\textsuperscript{b}] $\bar{X}$ and $\bar{Y}$ are the mean values of $X$ and $Y$, respectively. $X_{\tau}$ is $X$ at lag $\tau$. $\rho_{XX_{\tau}}$ and $\rho_{YY_{\tau}}$ are the auto-correlations (ACF) or partial autocorrelation (PACF) at lag $\tau$ of $X$ and $Y$ respectively. For the cross-correlation, we used $\tau_{max} = \lfloor 10 \times log(T/2) \rfloor$. For ACF and PACF, we used $\tau_{max} = 50$.
  \end{flushleft}
\end{table}

\begin{table}[!ht]
  \centering
  \footnotesize
  \caption{Time series distance functions used in this work} 
  \label{tab:measures_definitions_3}
  \begin{tabular}[t]{llc}
    \toprule
    Distance & Equation  & Ref. \\
    \midrule        
    
    \multirow{3}{*}{\pbox{22cm}{Temporal\\Correlation\\and Raw\\Values \textsuperscript{d}\\(CORT)}} &
      $ d_{\text{CORT}}(X,Y) = \displaystyle \Phi(cort(X,Y)) \cdot d_{\text{ED}}(X,Y) $ & 
      \cite{chouakria07}\\ [10px]
      & $\Phi(u) = \frac{2}{1+e^{ku}}$ &\\ [10px]
      & $cort(X,Y) = \displaystyle \frac{\sum_{i=1}^{T-1}(X_{t+1}-X_t)(Y_{t+1}-Y_t)}{\sqrt{\sum_{i=1}^{T-1}(X_{t+1}-X_t)^2}\sqrt{\sum_{i=1}^{T-1}(Y_{t+1}-Y_t)^2}}$ &\\ [22px]

    \rowcolor{gray!7}
    \pbox{22cm}{Short Time\\Series\textsuperscript{c} (STS)} &
      $d_{\text{STS}}(X,Y) = \sqrt{\sum_{t=1}^{T-1}\big((Y_{t+1}-Y_t)-(X_{t+1}-X_t)\big)^2}$ &
      \cite{moller03}\\ [15px]

    DISSIM &
      $d_{\text{DISSIM}}(X,Y) = \displaystyle \sum_{t=1}^{T-1} \int_t^{t+1} |X_t - Y_t| \mathrm{d}t$ &
      \cite{frentzos07}\\ [15px]   

    \rowcolor{gray!7}
    \pbox{22cm}{Normalized\\Compression\textsuperscript{e}\\(NCD)} &
    $ d_{\text{NCD}}(X,Y) = \displaystyle \frac{C(XY)-\min(C(X),C(Y))}{\max(C(X),C(Y))} $ & 
      \cite{cilibrasi05} \\ [8px]

    \pbox{22cm}{Compression-\\based\textsuperscript{d}\\(CDM)} & 
    $d_{CDM}(X,Y) = \displaystyle \frac{C(X,Y)}{C(X)+C(Y)}$
    & \cite{keogh07} \\ [8px]
      
    \rowcolor{gray!7}
    \pbox{22cm}{Periodogram\\(PER)} &
      $d_{\text{PER}}(X,Y) = \displaystyle \sqrt{\sum_{k=1}^{T/2}\big(I_X(\lambda_k)-I_Y(\lambda_k)\big)^2}$ & \cite{caiado06} \\ [22px]
      \rowcolor{gray!7}
      &$\lambda_{k} = \frac{2 \pi k}{T}$&\\ [8px]
      \rowcolor{gray!7}
      &$I_X(\lambda_k) = \frac{1}{T}\Big|\sum_{t=1}^{T}X_te^{-i\lambda_{k}t}\Big|^2$&\\ [8px]

    \pbox{22cm}{Integrated\\Periodogram\\(INTPER)} &
      $ d_{\text{INTPER}}(X,Y) = \displaystyle \int_{-\pi}^{\pi}|F_X(\lambda) - F_Y(\lambda)| \mathrm{d}\lambda$ & \cite{casado03} \\
      & $F_X(\lambda_j) = \displaystyle \frac{\sum_{i=1}^{j}I_X(\lambda_i)}{\sum_{i}I_X(\lambda_i)}
      $ &  \\ [15px]

    \rowcolor{gray!7}
    \pbox{22cm}{Fourier\\Coefficient\textsuperscript{f}\\(fourierDist)}  & 
      $ d_{FO}(X,Y) = \displaystyle \sqrt{\sum_{j=1}^{n}|x_j - y_j|^2}$ & \cite{agrawal93} \\ 
      \rowcolor{gray!7}
      &$x_j = \displaystyle \sum_{k=1}^T X_k \cdot e^{2\pi i (k-1)(j-1)/T}$&\\

    \bottomrule
  \end{tabular}
  \begin{flushleft} 
        \item $X$ and $Y$ are two time series with the same length $T$.
        \item[\textsuperscript{c}]The STS distance considers that the differences between two time series is the difference between all the slopes (linear functions) between the consecutive values in the time series. In our case, all the time series are equally spaced ($t=1$) with the same sampling rate. Therefore, the slope is $X_{i+1}-X_i$; 
        \item{\textsuperscript{d} $\Phi(u)$ and $k$ control the weight that $cort(X,Y)$ has on $d_{\text{ED}}(X,Y)$. We used $k=2$.}
        \item{\textsuperscript{e}} $C(X)$ is a compression function that returns the length of the compressed time series $X$. $C(X,Y)$ is the compressed size of the concatenation of $X$ and $Y$. We used three compressors: \texttt{gzip}, \texttt{bzip2} and \texttt{xz}, and returned the best (min) of them.
        \item[\textsuperscript{f}] $x_j$ and $y_j$ are the j-$th$ coefficients of the discrete Fourier transforms of $X$ and $Y$ respectively. $i=\sqrt{-1}$ is the imaginary unit We used the first $n=\lfloor T/2 \rfloor + 1$ coefficients.
  \end{flushleft}
\end{table}


\clearpage

\section{Climate network characteristics} 
\label{appendix:climate_nets_measures}

We present in Tab.~\ref{tab:climate_net_measures_1}, \ref{tab:climate_net_measures_2}, \ref{tab:climate_net_measures_3} and \ref{tab:climate_net_measures_4} the values of all network measures obtained for the climate networks constructed for the four edge densities $p=$ 0.001, 0.01, 0.05 and 0.1.

\begin{table}[!ht]
\caption{Global measures from networks created with percentile $p=0.001$} 
\label{tab:climate_net_measures_1}
\normalsize
\centering
\begin{tabular}{rcccccc}
  \toprule
  \textit{distance} & $LC$ & $L$ & $C$ & SWI & $r$ & $Q$ \\ 
  \midrule
  \rowcolor{gray!7}
  acf & 25.00 \% & 39.48 & 0.93 & 1.82 & 0.97 & 0.78 \\ 
  avgL1LInf & 67.39 \% & 7.77 & 0.87 & 9.73 & 0.95 & 0.72 \\ 
  \rowcolor{gray!7}
  bhattacharyya & 68.79 \% & 6.80 & 0.84 & 10.98 & 0.92 & 0.71 \\ 
  cdm & 62.85 \% & 3.50 & 0.61 & 19.71 & 0.22 & 0.55 \\ 
  \rowcolor{gray!7}
  cid &  9.32 \% & 18.01 & 0.90 & 3.32 & 0.94 & 0.68 \\ 
  correlation & 63.54 \% & 105.45 & 0.92 & 0.66 & 0.95 & 0.76 \\ 
  \rowcolor{gray!7}
  cort & 21.63 \% & 53.40 & 0.90 & 1.12 & 0.93 & 0.67 \\ 
  crossCorr & 64.00 \% & 105.45 & 0.92 & 0.66 & 0.95 & 0.76 \\ 
  \rowcolor{gray!7}
  dice & 69.00 \% & 9.15 & 0.88 & 7.93 & 0.95 & 0.73 \\ 
  dissim & 37.27 \% & 29.64 & 0.87 & 2.30 & 0.93 & 0.75 \\ 
  \rowcolor{gray!7}
  dtw &  7.84 \% & 9.85 & 0.89 & 6.54 & 0.94 & 0.74 \\ 
  euclidean & 17.89 \% & 49.56 & 0.90 & 1.20 & 0.93 & 0.66 \\ 
  \rowcolor{gray!7}
  fourierDist &  8.45 \% & 16.38 & 0.90 & 3.61 & 0.92 & 0.66 \\ 
  gower & 65.84 \% & 8.13 & 0.89 & 9.00 & 0.96 & 0.73 \\ 
  \rowcolor{gray!7}
  intper & 28.30 \% & 35.15 & 0.95 & 1.96 & 0.98 & 0.81 \\ 
  jaccard & 69.00 \% & 9.15 & 0.88 & 7.97 & 0.95 & 0.73 \\ 
  \rowcolor{gray!7}
  kulczynski & 65.84 \% & 8.13 & 0.89 & 9.03 & 0.96 & 0.73 \\ 
  lorentzian & 69.05 \% & 7.97 & 0.87 & 9.24 & 0.95 & 0.72 \\ 
  \rowcolor{gray!7}
  manhattan &  9.08 \% & 20.09 & 0.89 & 2.94 & 0.92 & 0.66 \\ 
  mi & 55.31 \% & 125.36 & 0.90 & 0.53 & 0.95 & 0.75 \\ 
  \rowcolor{gray!7}
  mic & 69.82 \% & 70.53 & 0.94 & 1.05 & 0.98 & 0.81 \\ 
  ncd & 62.54 \% & 3.51 & 0.61 & 19.75 & 0.24 & 0.55 \\ 
  \rowcolor{gray!7}
  pacf & 34.04 \% & 43.74 & 0.91 & 1.55 & 0.96 & 0.74 \\ 
  per &  8.00 \% & 15.26 & 0.88 & 3.71 & 0.90 & 0.57 \\ 
  \rowcolor{gray!7}
  sorensen & 65.84 \% & 8.13 & 0.89 & 9.02 & 0.96 & 0.73 \\ 
  sqdEuclidean & 68.85 \% & 7.41 & 0.81 & 10.55 & 0.90 & 0.70 \\ 
  \rowcolor{gray!7}
  sts & 41.16 \% & 63.69 & 0.91 & 1.07 & 0.96 & 0.77 \\ 
  tanimoto & 65.30 \% & 8.17 & 0.90 & 9.02 & 0.96 & 0.73 \\ 
  \rowcolor{gray!7}
  waveHedges & 69.05 \% & 7.97 & 0.87 & 9.26 & 0.95 & 0.72 \\ 
  \bottomrule
\end{tabular}
\end{table}

\begin{table}[!ht]
\caption{Global measures from networks created with percentile $p=0.01$} 
\label{tab:climate_net_measures_2}
\normalsize
\centering
\begin{tabular}{rcccccc}
  \toprule
  \textit{distance} & $LC$ & $L$ & $C$ & SWI & $r$ & $Q$ \\ 
  \midrule
  \rowcolor{gray!7}
  acf & 100.00 \% & 6.10 & 0.56 & 8.06 & 0.73 & 0.48 \\ 
  avgL1LInf &  97.83 \% & 5.03 & 0.31 & 6.29 & 0.19 & 0.46 \\ 
  \rowcolor{gray!7}
  bhattacharyya &  96.39 \% & 4.55 & 0.31 & 6.51 & 0.17 & 0.44 \\ 
  cdm &  97.75 \% & 2.48 & 0.11 & 4.36 & -0.14 & 0.24 \\ 
  \rowcolor{gray!7}
  cid & 100.00 \% & 6.45 & 0.76 & 6.90 & 0.87 & 0.66 \\ 
  correlation & 100.00 \% & 10.76 & 0.76 & 4.40 & 0.88 & 0.64 \\ 
  \rowcolor{gray!7}
  cort & 100.00 \% & 11.48 & 0.77 & 3.97 & 0.91 & 0.63 \\ 
  crossCorr & 100.00 \% & 10.74 & 0.76 & 4.41 & 0.88 & 0.64 \\ 
  \rowcolor{gray!7}
  dice &  98.00 \% & 4.92 & 0.35 & 7.13 & 0.20 & 0.47 \\ 
  dissim &  99.97 \% & 4.72 & 0.48 & 8.27 & 0.69 & 0.67 \\ 
  \rowcolor{gray!7}
  dtw &  95.14 \% & 7.31 & 0.62 & 4.37 & 0.39 & 0.56 \\ 
  euclidean & 100.00 \% & 7.11 & 0.78 & 6.26 & 0.88 & 0.64 \\ 
  \rowcolor{gray!7}
  fourierDist & 100.00 \% & 6.16 & 0.77 & 7.27 & 0.87 & 0.71 \\ 
  gower &  98.39 \% & 5.07 & 0.31 & 6.66 & 0.19 & 0.51 \\ 
  \rowcolor{gray!7}
  intper &  99.49 \% & 11.17 & 0.56 & 4.91 & 0.75 & 0.57 \\ 
  jaccard &  98.00 \% & 4.92 & 0.35 & 7.13 & 0.20 & 0.47 \\ 
  \rowcolor{gray!7}
  kulczynski &  98.39 \% & 5.07 & 0.31 & 6.66 & 0.19 & 0.51 \\ 
  lorentzian &  98.55 \% & 5.01 & 0.31 & 6.45 & 0.19 & 0.46 \\ 
  \rowcolor{gray!7}
  manhattan & 100.00 \% & 6.58 & 0.77 & 6.71 & 0.88 & 0.64 \\ 
  mi & 100.00 \% & 10.23 & 0.76 & 4.44 & 0.89 & 0.65 \\ 
  \rowcolor{gray!7}
  mic & 100.00 \% & 9.61 & 0.75 & 5.01 & 0.86 & 0.63 \\ 
  ncd &  97.62 \% & 2.49 & 0.11 & 4.37 & -0.13 & 0.24 \\ 
  \rowcolor{gray!7}
  pacf &  99.94 \% & 4.12 & 0.65 & 11.08 & 0.81 & 0.59 \\ 
  per &  94.00 \% & 9.40 & 0.68 & 2.81 & 0.51 & 0.32 \\ 
  \rowcolor{gray!7}
  sorensen &  98.39 \% & 5.07 & 0.31 & 6.66 & 0.19 & 0.51 \\ 
  sqdEuclidean &  94.34 \% & 4.80 & 0.32 & 6.18 & 0.20 & 0.47 \\ 
  \rowcolor{gray!7}
  sts &  99.99 \% & 8.43 & 0.73 & 3.90 & 0.59 & 0.58 \\ 
  tanimoto &  98.35 \% & 5.09 & 0.31 & 6.71 & 0.19 & 0.51 \\ 
  \rowcolor{gray!7}
  waveHedges &  98.55 \% & 5.01 & 0.31 & 6.45 & 0.19 & 0.46 \\ 
  \bottomrule
\end{tabular}
\end{table}

\begin{table}[!ht]
\caption{Global measures from networks created with percentile $p=0.05$} 
\label{tab:climate_net_measures_3}
\normalsize
\centering
\begin{tabular}{rcccccc}
  \toprule
  \textit{distance} & $LC$ & $L$ & $C$ & SWI & $r$ & $Q$ \\ 
  \midrule
  \rowcolor{gray!7}
  acf & 100.00 \% & 4.17 & 0.56 & 3.53 & 0.59 & 0.43 \\ 
  avgL1LInf &  99.82 \% & 3.72 & 0.46 & 3.36 & 0.32 & 0.39 \\ 
  \rowcolor{gray!7}
  bhattacharyya &  99.77 \% & 3.47 & 0.46 & 3.51 & 0.30 & 0.40 \\ 
  cdm &  99.94 \% & 2.03 & 0.21 & 2.65 & -0.16 & 0.18 \\ 
  \rowcolor{gray!7}
  cid & 100.00 \% & 2.64 & 0.53 & 5.03 & 0.46 & 0.40 \\ 
  correlation & 100.00 \% & 3.06 & 0.65 & 5.32 & 0.45 & 0.56 \\ 
  \rowcolor{gray!7}
  cort & 100.00 \% & 2.50 & 0.61 & 6.86 & 0.64 & 0.52 \\ 
  crossCorr & 100.00 \% & 3.02 & 0.62 & 5.00 & 0.42 & 0.55 \\ 
  \rowcolor{gray!7}
  dice & 100.00 \% & 3.58 & 0.47 & 3.58 & 0.30 & 0.41 \\ 
  dissim & 100.00 \% & 2.54 & 0.44 & 4.34 & 0.25 & 0.32 \\ 
  \rowcolor{gray!7}
  dtw &  99.48 \% & 3.33 & 0.58 & 3.52 & 0.09 & 0.33 \\ 
  euclidean & 100.00 \% & 2.47 & 0.51 & 5.20 & 0.25 & 0.42 \\ 
  \rowcolor{gray!7}
  fourierDist & 100.00 \% & 2.50 & 0.49 & 5.04 & 0.23 & 0.40 \\ 
  gower &  99.88 \% & 3.66 & 0.46 & 3.47 & 0.33 & 0.40 \\ 
  \rowcolor{gray!7}
  intper &  99.95 \% & 6.74 & 0.60 & 2.66 & 0.64 & 0.55 \\ 
  jaccard & 100.00 \% & 3.58 & 0.47 & 3.58 & 0.30 & 0.41 \\ 
  \rowcolor{gray!7}
  kulczynski &  99.88 \% & 3.66 & 0.46 & 3.47 & 0.33 & 0.40 \\ 
  lorentzian &  99.89 \% & 3.63 & 0.46 & 3.45 & 0.33 & 0.43 \\ 
  \rowcolor{gray!7}
  manhattan & 100.00 \% & 2.53 & 0.52 & 5.19 & 0.28 & 0.41 \\ 
  mi & 100.00 \% & 2.26 & 0.53 & 6.17 & 0.41 & 0.53 \\ 
  \rowcolor{gray!7}
  mic & 100.00 \% & 2.84 & 0.63 & 5.38 & 0.40 & 0.56 \\ 
  ncd &  99.93 \% & 2.04 & 0.21 & 2.65 & -0.16 & 0.18 \\ 
  \rowcolor{gray!7}
  pacf &  99.99 \% & 2.67 & 0.41 & 4.32 & 0.17 & 0.43 \\ 
  per &  99.00 \% & 4.58 & 0.76 & 2.80 & 0.26 & 0.12 \\ 
  \rowcolor{gray!7}
  sorensen &  99.88 \% & 3.66 & 0.46 & 3.47 & 0.33 & 0.40 \\ 
  sqdEuclidean &  98.49 \% & 3.81 & 0.48 & 3.22 & 0.31 & 0.39 \\ 
  \rowcolor{gray!7}
  sts & 100.00 \% & 3.37 & 0.69 & 3.73 & 0.18 & 0.27 \\ 
  tanimoto &  99.88 \% & 3.66 & 0.46 & 3.47 & 0.33 & 0.40 \\ 
  \rowcolor{gray!7}
  waveHedges &  99.89 \% & 3.63 & 0.46 & 3.45 & 0.33 & 0.43 \\ 
  \bottomrule
\end{tabular}
\end{table}

\begin{table}[!ht]
\caption{Global measures from networks created with percentile $p=0.1$} 
\label{tab:climate_net_measures_4}
\normalsize
\centering
\begin{tabular}{rcccccc}
  \toprule
  \textit{distance} & $LC$ & $L$ & $C$ & SWI & $r$ & $Q$ \\ 
  \midrule
  \rowcolor{gray!7}
  acf & 100.00 \% & 3.44 & 0.64 & 2.71 & 0.61 & 0.35 \\ 
  avgL1LInf &  99.92 \% & 3.17 & 0.57 & 2.73 & 0.43 & 0.37 \\ 
  \rowcolor{gray!7}
  bhattacharyya &  99.97 \% & 2.98 & 0.56 & 2.83 & 0.40 & 0.36 \\ 
  cdm &  99.99 \% & 1.93 & 0.29 & 2.20 & -0.16 & 0.16 \\ 
  \rowcolor{gray!7}
  cid & 100.00 \% & 2.27 & 0.53 & 3.41 & 0.32 & 0.29 \\ 
  correlation & 100.00 \% & 2.31 & 0.58 & 3.64 & 0.29 & 0.41 \\ 
  \rowcolor{gray!7}
  cort & 100.00 \% & 2.00 & 0.46 & 3.67 & 0.23 & 0.33 \\ 
  crossCorr & 100.00 \% & 2.37 & 0.59 & 3.42 & 0.24 & 0.30 \\ 
  \rowcolor{gray!7}
  dice & 100.00 \% & 3.08 & 0.57 & 2.80 & 0.41 & 0.39 \\ 
  dissim & 100.00 \% & 2.17 & 0.48 & 3.16 & 0.08 & 0.28 \\ 
  \rowcolor{gray!7}
  dtw &  99.88 \% & 2.58 & 0.61 & 3.02 & -0.01 & 0.23 \\ 
  euclidean & 100.00 \% & 2.09 & 0.49 & 3.37 & 0.04 & 0.31 \\ 
  \rowcolor{gray!7}
  fourierDist & 100.00 \% & 2.16 & 0.49 & 3.32 & 0.05 & 0.30 \\ 
  gower &  99.90 \% & 3.14 & 0.56 & 2.80 & 0.43 & 0.37 \\ 
  \rowcolor{gray!7}
  intper & 100.00 \% & 5.12 & 0.67 & 2.12 & 0.70 & 0.48 \\ 
  jaccard & 100.00 \% & 3.08 & 0.57 & 2.80 & 0.41 & 0.39 \\ 
  \rowcolor{gray!7}
  kulczynski &  99.90 \% & 3.14 & 0.56 & 2.80 & 0.43 & 0.37 \\ 
  lorentzian &  99.90 \% & 3.11 & 0.57 & 2.77 & 0.44 & 0.37 \\ 
  \rowcolor{gray!7}
  manhattan & 100.00 \% & 2.13 & 0.51 & 3.41 & 0.07 & 0.33 \\ 
  mi & 100.00 \% & 1.90 & 0.42 & 3.45 & 0.26 & 0.35 \\ 
  \rowcolor{gray!7}
  mic & 100.00 \% & 2.07 & 0.55 & 3.81 & 0.22 & 0.38 \\ 
  ncd &  99.99 \% & 1.94 & 0.30 & 2.20 & -0.16 & 0.16 \\ 
  \rowcolor{gray!7}
  pacf & 100.00 \% & 2.35 & 0.46 & 3.06 & 0.20 & 0.33 \\ 
  per & 100.00 \% & 3.41 & 0.80 & 2.71 & 0.24 & 0.09 \\ 
  \rowcolor{gray!7}
  sorensen &  99.90 \% & 3.14 & 0.56 & 2.80 & 0.43 & 0.37 \\ 
  sqdEuclidean &  99.41 \% & 3.31 & 0.58 & 2.57 & 0.42 & 0.36 \\ 
  \rowcolor{gray!7}
  sts & 100.00 \% & 2.56 & 0.62 & 2.93 & -0.17 & 0.19 \\ 
  tanimoto &  99.90 \% & 3.14 & 0.56 & 2.80 & 0.43 & 0.37 \\ 
  \rowcolor{gray!7}
  waveHedges &  99.90 \% & 3.11 & 0.57 & 2.77 & 0.44 & 0.37 \\ 
  \bottomrule
\end{tabular}
\end{table}

\clearpage
\section{Network distance matrices} 
\label{appendix:dnet_distances}

\begin{figure}[!htbp]
  \centering
  \subfloat{\label{fig:net_dist_a} \includegraphics[width=1\linewidth]{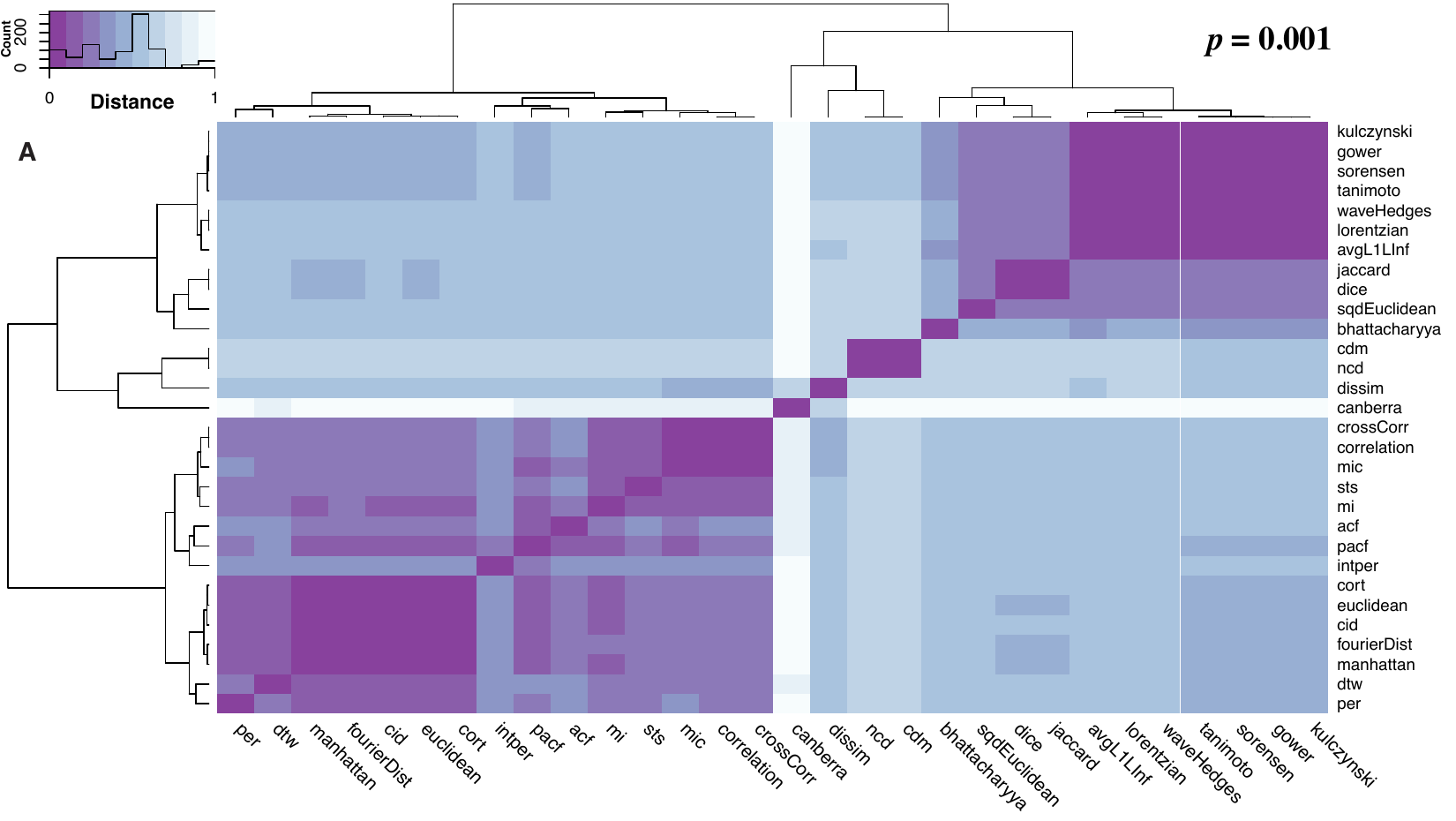}} \\
  \subfloat{\label{fig:net_dist_b} \includegraphics[width=1\linewidth]{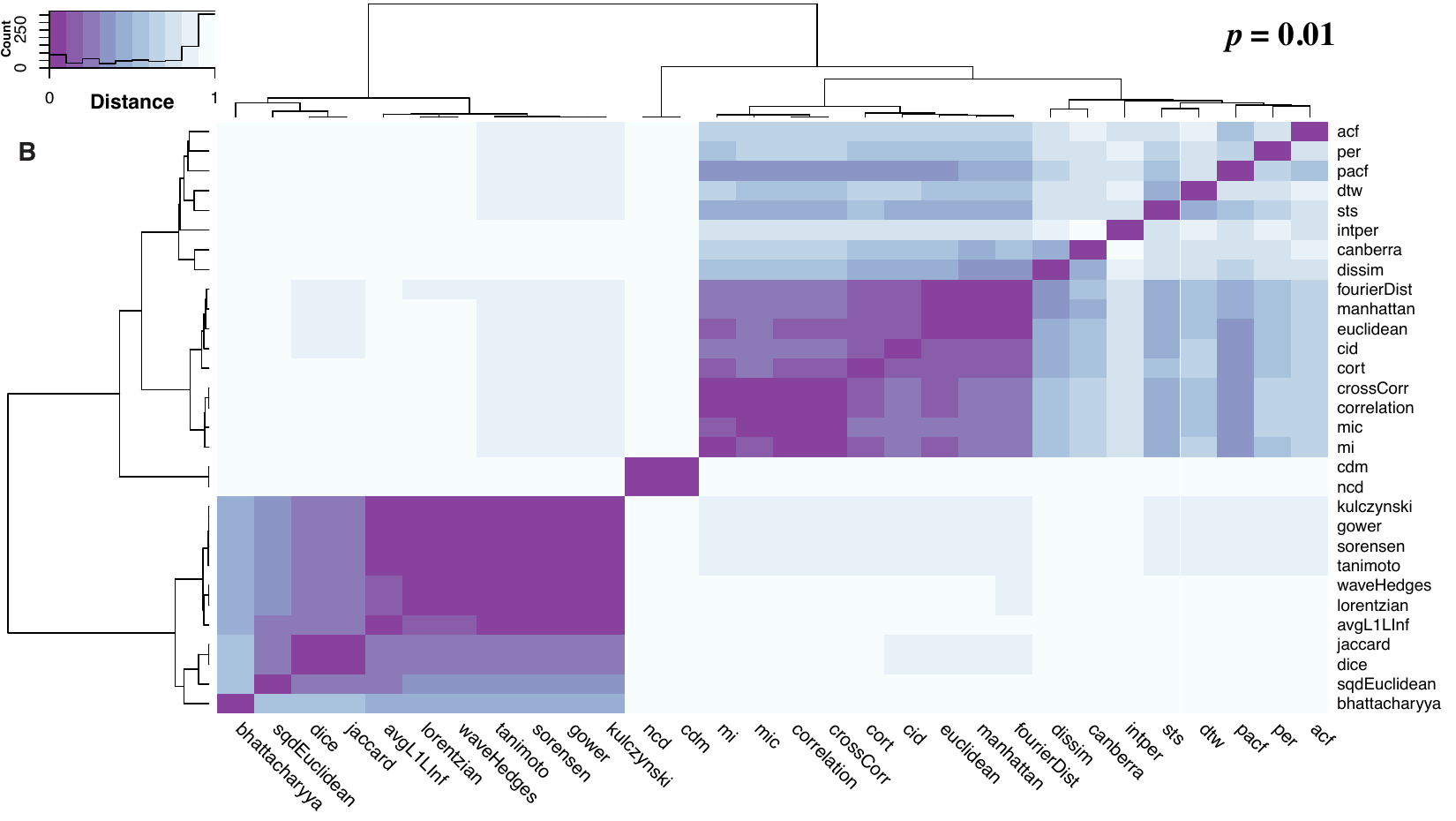}} 
\end{figure}

\begin{figure}[!htbp]
  \centering
  \subfloat{\label{fig:net_dist_c} \includegraphics[width=1\linewidth]{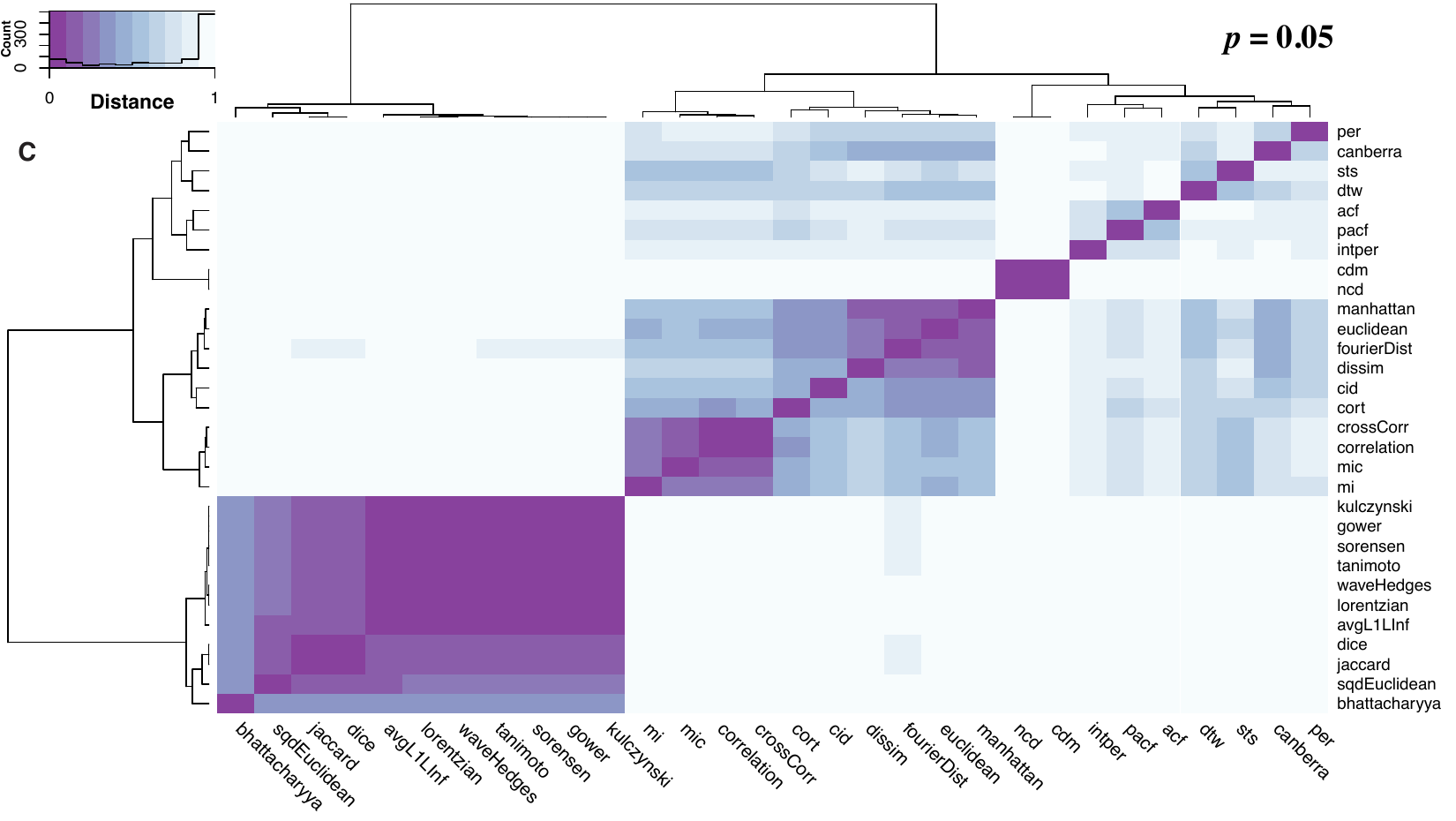}} \\
  \subfloat{\label{fig:net_dist_d} \includegraphics[width=1\linewidth]{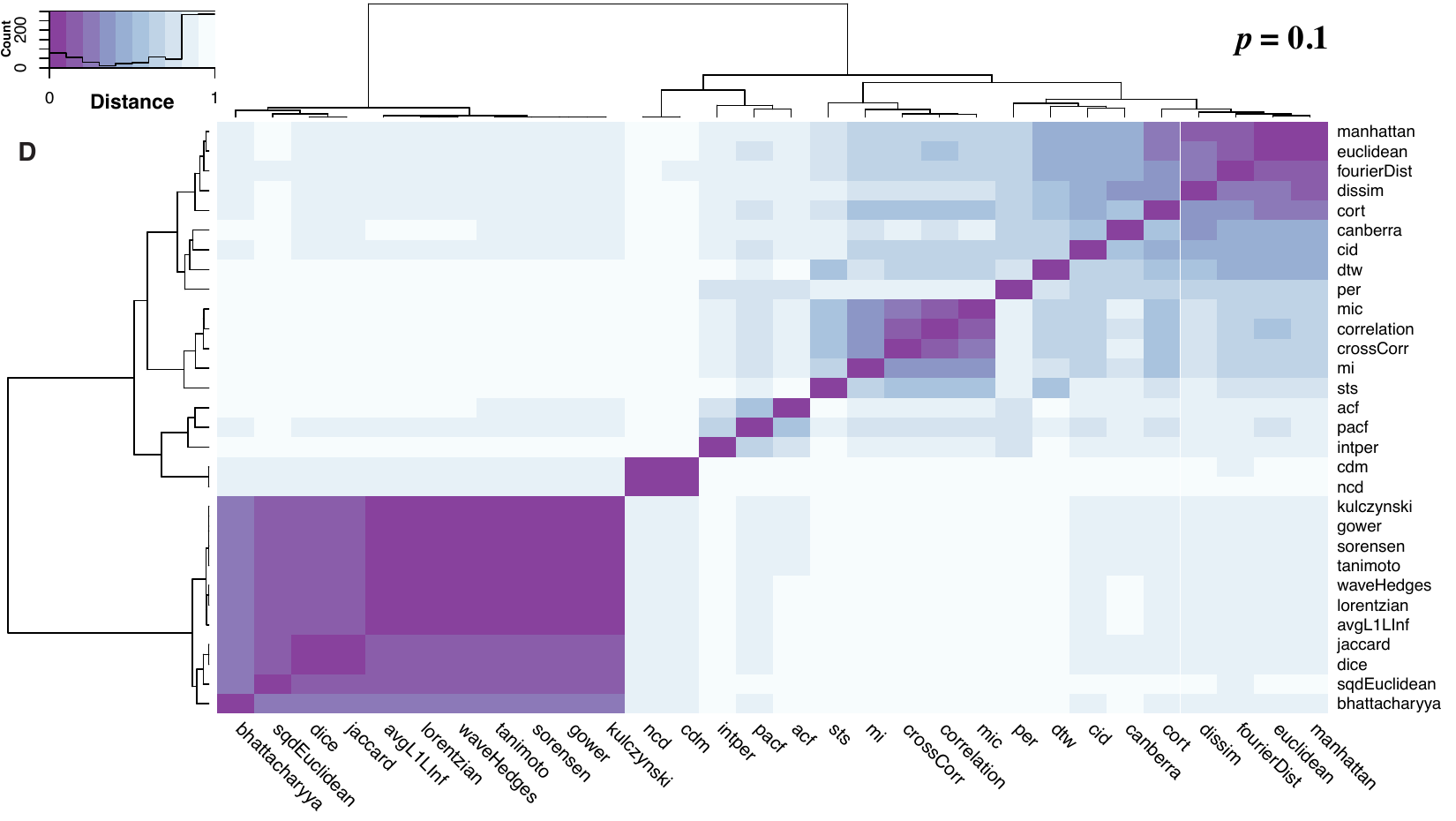}} 
  \caption{Network clustering according to the similarity of the respective network topology. For every distance function, we constructed climate networks considering four edge densities: $p=$ 0.001, 0.01, 0.05 and 0.1. Then, we calculated the Hamming distances between all pairs of networks and clustered them using hierarchical clustering (Ward's method). The results show that PDF-based distance functions form a cluster and generate similar networks independent of $p$. For the other measures, when $p$ increases, the networks were very different except in two small groups. The first group is formed by correlation, cross-correlation, MI and MIC. The other group is formed by Manhattan, Euclidean and Fourier distances.}
  \label{fig:net_dists} 
\end{figure}

\clearpage
\section{Teleconnections} 
\label{append:teleconnections}

\begin{figure}[!h]
  \centering
  \includegraphics[width=0.95\linewidth]{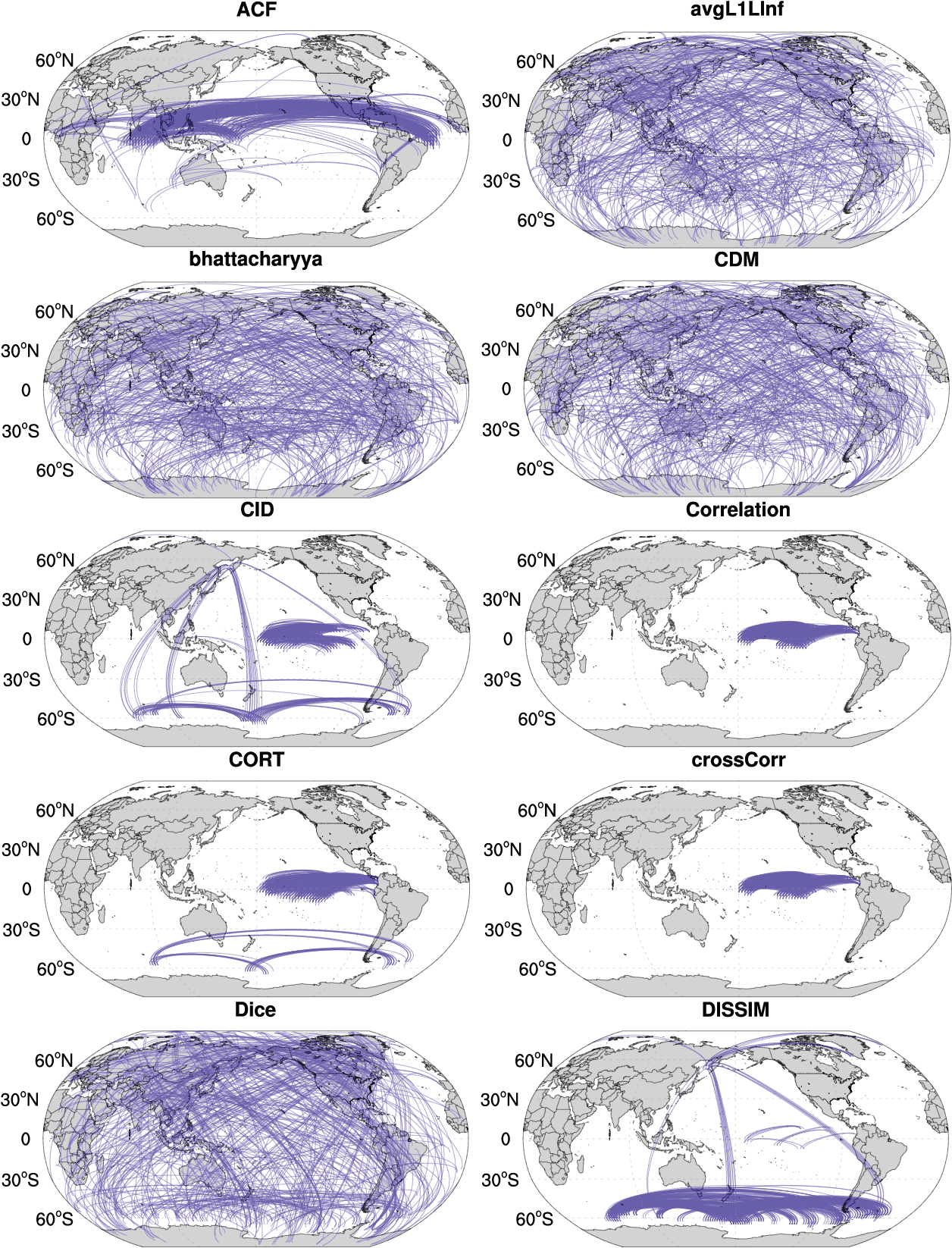} 
\end{figure}

\begin{figure}[!h]
  \centering
  \includegraphics[width=0.96\linewidth]{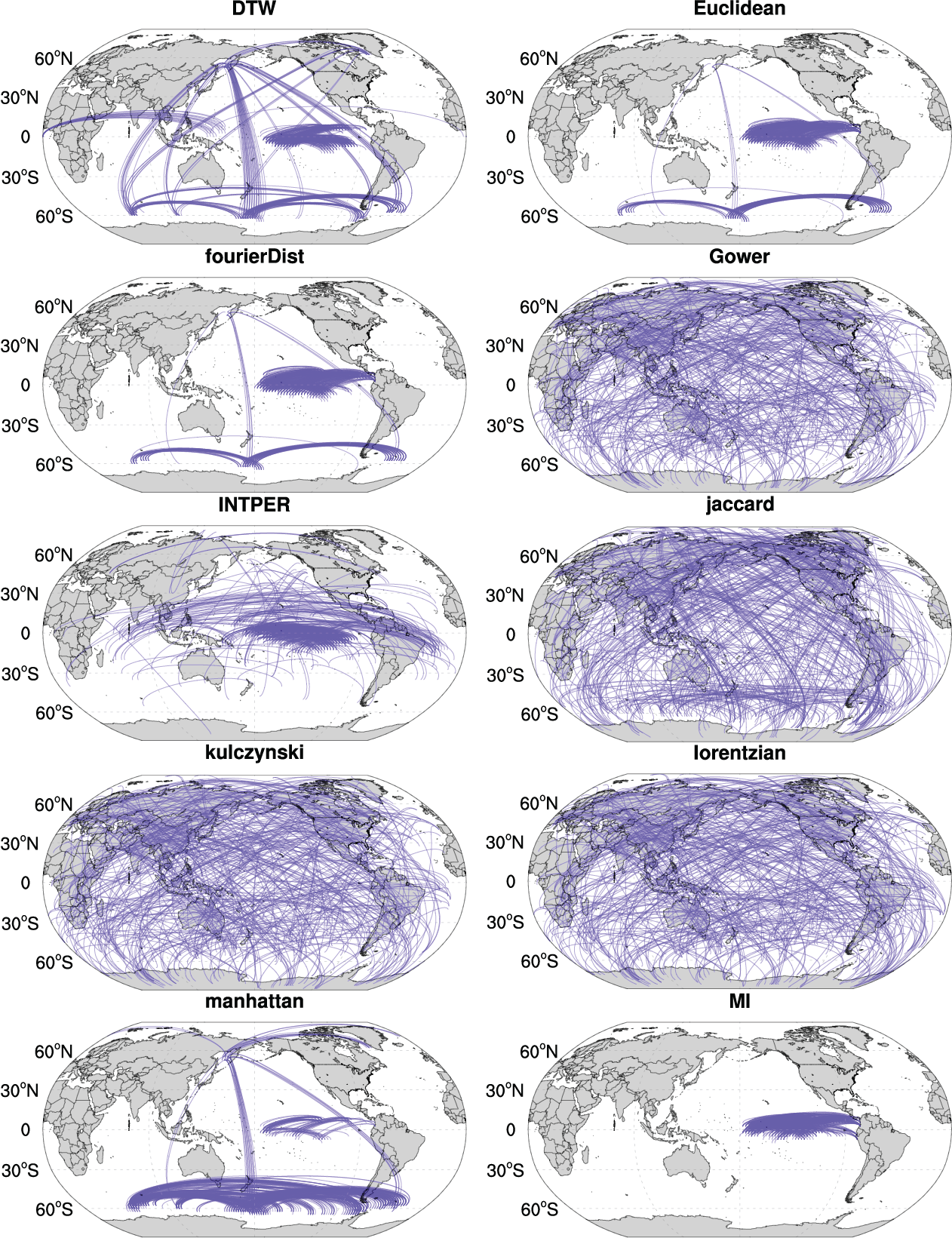} 
\end{figure}

\begin{figure}[!h]
  \centering
  \includegraphics[width=0.93\linewidth]{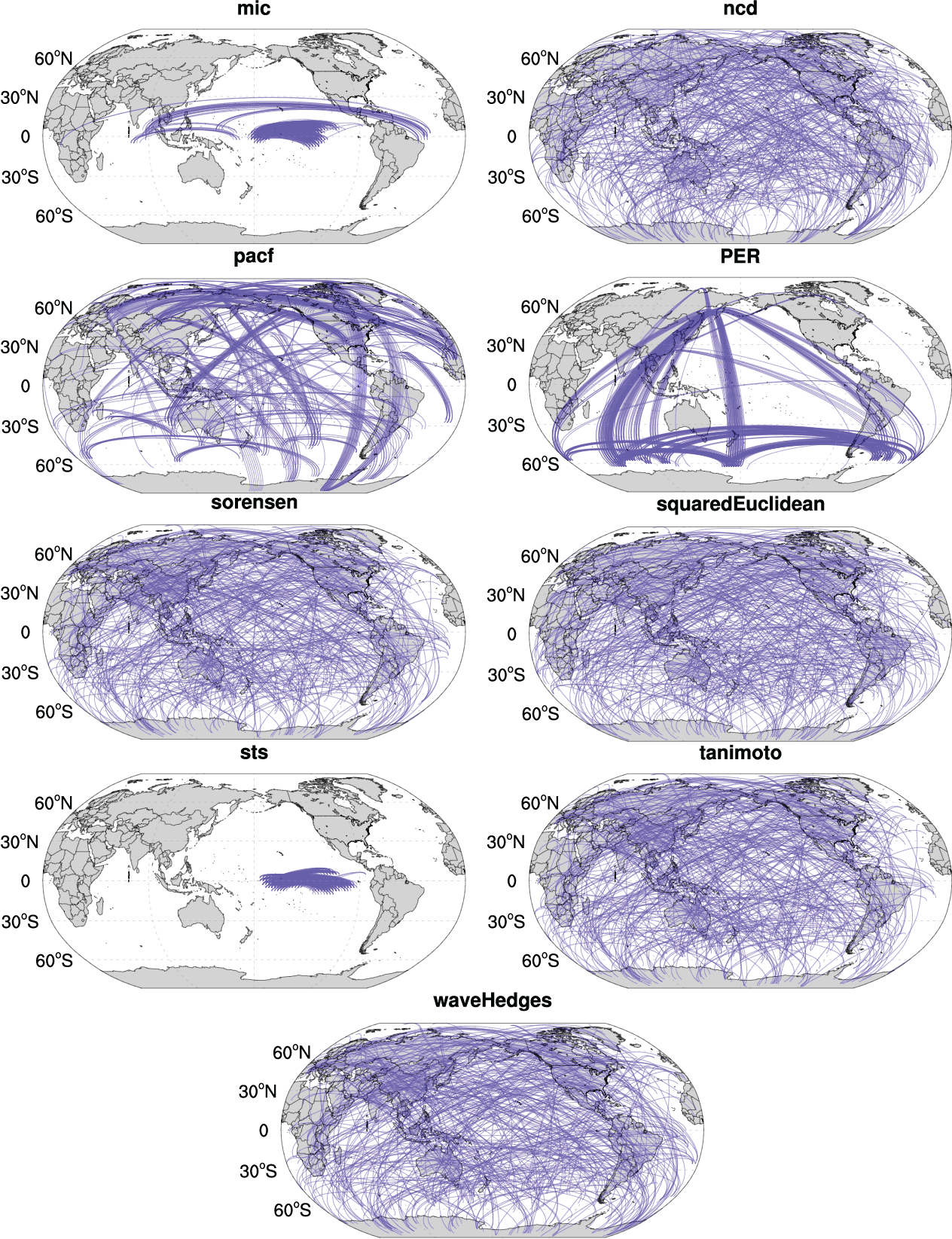} 
  \caption{Teleconnections for all the 29 distance functions. We considered only the 500 strongest teleconnections (nodes distance greater than $5000km$).}
  \label{fig:teleconnections_all} 
\end{figure}




\end{document}